\documentclass[journal,compsoc]{IEEEtran}

\usepackage{xcolor}              \usepackage[numbers]{natbib}             
\usepackage[hidelinks]{hyperref} \usepackage{xurl}                 \usepackage[inline]{enumitem}    \usepackage{syntax}              \usepackage{siunitx}             \DeclareSIUnit{\rpm}{rpm}
\usepackage{multirow}            \usepackage{makecell}
\usepackage{booktabs}
\usepackage{array}               \usepackage{subcaption}
\usepackage{amsmath,amsfonts}    \usepackage{amssymb}             \usepackage{newtxmath}
\usepackage{subcaption}
\usepackage{graphicx}

\usepackage{algorithm} \usepackage{algpseudocode}

\newcommand{\funof}[1]{(#1)}

\newcommand{\mi}[0]{\mu}

\definecolor{ref}{RGB}{0,0,0} \definecolor{exp}{RGB}{0,128,255} \definecolor{act}{RGB}{255,51,253}

\newcommand{\asv}[0]{$\alpha$}
\newcommand{\tsv}[0]{$\delta$}
\newcommand{\trc}[0]{$r$}

\newcommand{\SP}[0]{$\oplus$}
\newcommand{\AS}[0]{$\bullet$}
\newcommand{\TS}[0]{$\triangleright$}

\newcommand{\fitness}[0]{\mathcal{F}}
\newcommand{\ctrlerr}[0]{\mathcal{CE}}
\newcommand{\ctrlerrth}[0]{\mathcal{CE}_{\mathit{th}}}
\newcommand{\mrfalsify}[0]{\mathcal{MF}}
 \title{Testing CPS with Design Assumptions-Based Metamorphic Relations and Genetic Programming}

\author{Claudio~Mandrioli,
Seung~Yeob~Shin,~\IEEEmembership{Member,~IEEE},
Domenico~Bianculli, ~\IEEEmembership{Member,~IEEE},
and Lionel~Briand,~\IEEEmembership{Fellow,~IEEE}
\IEEEcompsocitemizethanks{
\IEEEcompsocthanksitem{C.\ Mandrioli, S.\ Y.\ Shin, and D.\ Bianculli are with the University of Luxembourg, Luxembourg.}
\IEEEcompsocthanksitem{L.\ Briand is with the University of Ottawa, Canada, and the Research Ireland Lero centre for software, University of
Limerick, Ireland.}
}
}

\begin{document}

\maketitle
\begin{abstract}
Cyber-Physical Systems (CPSs) software is used to enforce desired behaviours on physical systems.
To test the interaction between the CPS software and the system's physics, engineers provide traces of desired physical states and observe traces of the actual physical states.
CPS requirements describe how closely the actual physical traces should track the desired traces.
These requirements are typically defined for specific, simple input traces such as step or ramp sequences, and thus are not applicable to arbitrary inputs.
This limits the availability of oracles for CPSs.
Our recent work proposes an approach to testing CPSs using control-theoretical design assumptions instead of requirements.
This approach circumvents the oracle problem by leveraging the control-theoretical guarantees that are provided when the design assumptions are satisfied.
To address the test case generation and oracle problems, researchers have proposed metamorphic testing, which is based on the study of relations across tests, i.e., metamorphic relations (MRs).

In this work, we define MRs based on the design assumptions and explore combinations of these MRs using genetic programming to generate CPS test cases.
This enables the generation of CPS input traces with potentially arbitrary shapes, together with associated expected output traces.
We use the deviation from the expected output traces to guide the generation of input traces that falsify the MRs.
Our experiment results show that the MR-falsification provides engineers with new information, helping them identify passed and failed test cases.
Furthermore, we show that the generation of traces that falsify the MRs is a non-trivial problem, which cannot be addressed with a random generation approach but is successfully addressed by our approach based on genetic search.
\end{abstract} 
\section{Introduction}
\label{sec:introduction}
Cyber-Physical Systems (CPSs) are pervasive in modern life.
Such systems are characterised by the closed-loop interaction between software and physical parts~\cite{lee:2015}.
The goal of this interaction is to have the software enforce a desired behaviour on the physical part.
More precisely, the software should steer the physical part to desired states, for example flying a drone to the desired position.
Accordingly, CPSs requirements specify the desired behaviour of the physical part rather than the software behaviour.
This makes the testing of CPSs software a multidisciplinary problem (including aspects of software and control engineering), regarding both the generation of test inputs and the implementation of oracles~\cite{mandrioli:2023b}.

\begin{figure}[t]
    \centering
    \includegraphics{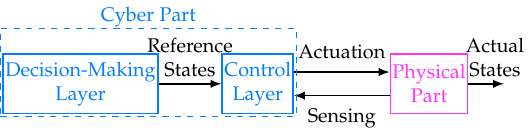}
    \caption{Structure of a CPS. The cyber part is represented by the dashed azure box; the physical part is represented by the solid purple box; arrows denote input or output signals.
    }
    \label{fig:cps-structure}
\end{figure}

Figure~\ref{fig:cps-structure} shows the structure of a CPS;  with the cyber part enclosed by the dashed azure box and the physical part enclosed by the solid purple box; arrows denote input or output signals.
Within the cyber part, the control layer is responsible for the real-time interaction with the physical part, performed through actuators and sensors.
The CPS decision-making layer defines a desired physical state (called {\em reference} in control-engineering jargon) that the control layer should achieve.
The control layer uses sensors and actuators to steer the physical part to the desired state.
For example, in a patrolling drone, the decision-making layer decides which areas to target and the path to follow.
Then, it sends the sequence of desired positions to the control layer, which is responsible for actually flying the drone through them.
Specifically, the control layer uses sensors to identify the drone position and sends commands to actuators to steer it to the desired position.

Software and control engineers collaboratively develop the control layer.
Control engineers design control algorithms (like PIDs and state-feedback controllers) in the form of equations.
Such algorithms are standardised; CPS development platforms like Matlab\footnote{
    \url{https://nl.mathworks.com/products/matlab.html}
}
provide graphical interfaces and automatic code generation tools for them.
However, their implementation on embedded devices, their interactions (even a small drone has a dozen of different PIDs interconnected in different ways), and their integration with other software components make their implementation non-trivial.
Thus, the implementation of control algorithms also requires software engineering expertise.
As such, {\em faults can appear from the control design, the software implementation, or their interactions}, thus calling for an interdisciplinary approach to reveal them.
For example, if there is a fault in the implementation of the positioning algorithm, a very aggressive tuning of a control algorithm can make the CPS more reactive to position changes and more sensitive to the fault; conversely, a less reactive tuning can make the drone more robust.

When testing the software implementing the CPS control layer, engineers define input traces of physical states reference values and observe output traces of the actual states.
Intuitively, the requirements define how closely the actual physical state should follow the reference {\em depending on a given input shape}.
Examples of typical requirements are the steady-state error, which  prescribes the maximum allowed difference between the desired and actual state when the input is constant, and the overshoot, which defines the maximum state value allowed after a step increase in the reference (i.e., an instant increase followed by constant values)~\cite{Astrom:2008}.
Such requirements can then be used to assess the CPS for constant inputs and step changes.
In practice, this tight connection between requirements and a given input trace limits the possibility of defining oracles applicable to different traces.
Indeed, a CPS is expected to track a variety of input traces, and possibly arbitrary ones.
Thus, most of the possible input traces are different from the ones for which the requirements are defined; it is then impossible to define oracles for them.

Currently, to overcome this limitation, engineers can use the distance between the desired and actual states, i.e., the {\em control error}, to assess the CPSs ability to track the reference states.
However, the control error is a coarse metric; it does not capture properties related to the shape of the input and output traces.
Thus, it has limited use as an oracle.

As an alternative to testing CPSs based on requirements, in our previous work~\cite{mandrioli:2023b} we proposed a new approach based on control theoretical guarantees.
The idea of the approach is to focus on testing scenarios that falsify the {\em design assumptions} that underlie the CPS mathematical models used by control engineers.
In fact, using  such mathematical models of a CPS behaviour, control engineers can provide a priori guarantees on the performance of the implemented CPS.
Then, when the design assumptions hold in the CPS implementation, we can rely on these guarantees from the control theoretical models for satisfying requirements.
In contrast, when the assumptions are falsified, the CPS behaviour is not predicted by the models and therefore it requires empirical verification, i.e., testing.
Thus, by testing the design assumptions, we can indirectly obtain guarantees on the requirements satisfaction through the control theoretical guarantees.
In our previous work~\cite{mandrioli:2023b}, we focused on how to generate test cases to falsify the {\em linear behaviour} design assumption~\cite{Astrom:2008,Hespanha:09}, which is widely used in CPS development~\cite{Astrom:2008,Calanca:2016,Lopez:2023}.
However, our previous work is limited as it targets the testing of individual CPS states (e.g., only the altitude control in a drone) and requires user-defined periodic input shapes.

When testing CPSs, causing failures is not always difficult.
Analogously, the sole falsification of the linearity design assumption is not necessarily a difficult task.
Specifically, large and fast-changing inputs are likely to falsify the linearity assumption, as they can lead a CPS to an unexpected physical state~\cite{mandrioli:2023b}.
Such scenarios correspond to a high degree of falsification of the design assumptions, and can be considered {\em trivial failures}, as the CPS was never expected to perform such manoeuvres.
The more interesting and challenging problem is the generation and identification of \emph{subtle failures}, where a CPS shows subtle behaviour deviations~\cite{zhang:2024,Hildebrandt:2020}.
An example of a subtle deviation is an unexpected oscillation of a patrolling drone, which possibly does not significantly change the control error but can affect the ability of the drone to take pictures of its target.
Accordingly, when generating scenarios that falsify the design assumptions, we also want to avoid trivial failures.

Metamorphic Testing (MT) has been proposed in the software engineering literature as a way to address both the test case generation and the oracle problems~\cite{chen:2020}.
MT is based on Metamorphic Relations (MRs), which are properties defined over multiple inputs and outputs of two or more test cases.
Being defined over multiple test cases, MRs are used in MT to generate new follow-up test cases from initial ones, together with an associated expected output.
In practice, the applicability of MT is limited by the availability of MRs that often have to be manually defined~\cite{chen:2018}.
More in general, the automated generation of MRs is still an open problem~\cite{Ayerdi:2024,Ayerdi:2021}.
For example, in recent works that have applied MT to different CPSs, such as cars~\cite{Deng:2022,ayerdi:2023} and UAVs~\cite{Li:2021}, MRs are defined manually and are specific to the application scenario.

In this work, we target the testing of CPSs developed using traditional control theory and, more specifically, linear systems theory.
This includes common control algorithms like PIDs, state-feedback approaches, frequency-based approaches, and Linear-Quadratic regulators.
In such systems, we aim to falsify design assumptions and push the CPS at the boundaries of its acceptable performance.
In this way, we can expose faults in the software, the control design, or their interaction.

We propose to use the design assumptions to define MRs and apply MT to CPS testing.
We define MRs based on the control-theoretical design assumptions of linear behaviour~\cite{Hespanha:09}, which are independent of the specific application domain of the CPS under test (e.g., a drone or a car).
Using these MRs, we aim to test the assumption---on the control layer implementation---that such an implementation should behave linearly.
We express the linearity properties as MRs between input-output pairs of initial and follow-up tests.
Using the MRs, we generate input traces together with their associated expected output traces.
By comparing the expected output trace with the actual one, we provide engineers with a new metric, the \emph{MR-falsification degree}, which can be used to distinguish test cases within the design scope (i.e., those that follow their reference thanks to the control theoretical guarantees) from test cases that are out of the design scope and show unexpected behaviour.
Using a Genetic Programming (GP) algorithm, we then combine MR-falsification and the control error to guide the generation of input traces and their expected output traces, leading the CPS out of its design scope (i.e., falsifying the MRs) while avoiding trivial failures.

Our experiment results show that the problem of falsifying the MRs, while avoiding trivial failures, cannot be addressed with a random generation approach.
In contrast, the proposed GP approach can generate arbitrary input traces that falsify the MRs and thus the associated design assumptions, while avoiding trivial failures.
Furthermore, our results show that, for non-trivial failures, the MR-falsification degree is not correlated with the control error, and thus effectively provides new information for the assessment of CPS test outputs.

To summarise, this article makes the following contributions to the CPS test input generation and oracle problems:
\begin{itemize}
    \item the definition of MRs based on the linear behaviour design assumption, which are valid independently of the CPS application domain;
    \item the use of such MRs, in combination with GP and MT, to generate CPS input traces with arbitrary shapes and associated expected output traces.
    These traces lead to the falsification of the linearity design assumption, while avoiding trivial failures;
    \item the development of a novel approach for testing linearity design assumptions using multi-dimensional and non-periodic input traces, with potentially arbitrary shapes;
    \item the empirical evaluation of the proposed approach on three CPSs from the automotive, drone, and aircraft domains.
\end{itemize}

The remainder of the article is structured as follows.
Section~\ref{sec:motivating-example} exemplifies the challenges addressed in this work and defines our testing objective.
Section~\ref{sec:metamorphic-relations} introduces the MRs our approach relies upon.
Section~\ref{sec:approach} describes the proposed GP approach.
Section~\ref{sec:empirical-evaluation} reports on the empirical evaluation of our approach.
Section~\ref{sec:related-work} discusses the related work. 
Section~\ref{sec:conclusions} concludes the article and outlines future work.
 \begin{figure}[t]
    \centering
    \includegraphics{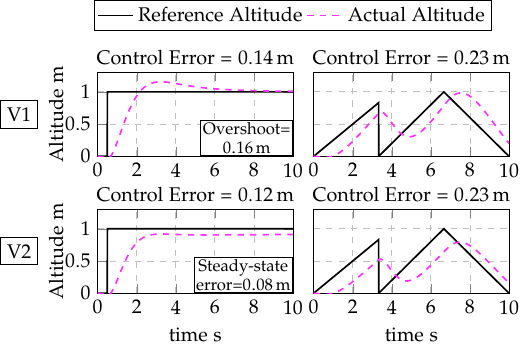}
    \caption{Examples of step responses of the altitude control of a drone, with the reference input altitude drawn using a solid black and the actual altitude drawn with a dashed purple line.
    The left-hand side plots show a step response, where the steady-state error and overshoot are well defined.
    The right-hand side plots show instead an arbitrary input trace, for which the steady state and overshoot cannot be assessed.
    }
    \label{fig:intro-examples}
\end{figure}

\section{Motivating Example and Testing Objective}
\label{sec:motivating-example}
In this section, we use the altitude control of a drone to exemplify the challenges addressed in this work.
Specifically, we showcase the limitations of requirements defined based on simple input shapes, and the limitations of the control error as an oracle.
We conclude the section defining our testing objective.

Figure~\ref{fig:intro-examples} shows two flight tests for two different implementations of a drone altitude control, with implementation V1 in the first row and V2 in the second row.
In the left-hand side plots, we show the test of a step input trace and, in the right-hand side, an arbitrary trace.
In each plot, the solid line represents the reference altitude (i.e., the input trace), and the purple dashed line shows the drone actual altitude (i.e., the test output); thus, the objective is that the actual altitude tracks the reference one.

In the step-input tests, we can directly assess the satisfaction of some requirements, such as the steady-state error and the overshoot.
In V1, we observe that the drone eventually reaches precisely the desired altitude when the input is constant, thus showing a zero steady-state error.
In contrast, V2 does not reach the desired value, and is thus not satisfying the steady-state error requirement.
The overshoot can be clearly measured at $t=3$ for V1, when the actual altitude reaches the maximum value.
Instead, for V2, the actual output never takes values larger than the desired one, so the overshoot is zero.
Thus, for the step input, we can directly check the steady-state error and overshoot requirements.

Differently, for the arbitrary input traces in the right-hand side plots, this assessment is not feasible.
First, the input is never constant, preventing the assessment of the steady-state error.
In addition, there is no clear time instant at which the overshoot should be assessed, since the input changes linearly rather than in steps.
As a result, such requirements are not well-defined for this arbitrary input trace, and oracles for assessing them cannot be implemented.

As mentioned in Section~\ref{sec:introduction}, engineers can always use the control error, i.e., the distance between the input reference and actual output, to assess the cases.
However, when using the control error to assess the tests in Figure~\ref{fig:intro-examples}, we observe that they show very similar values (indicated above the plots, and measured as the average difference between the desired and actual altitudes) for the two versions of the CPS.
Specifically, for the arbitrary input, we observe the very same control error value ($0.23$), and for the step input, we cannot explain the $0.14-0.12=0.02$ difference in control error between V1 and V2 by a different degree of satisfaction of the steady-state error and overshoot requirements.
This showcases why the control error, despite being the most intuitive metric for assessing a CPS ability to track references, is not sufficient to implement oracles.

To address the limitations of direct requirements assessment and of control error as CPSs oracles, we propose the use of design assumptions.
Most importantly, the design assumptions that underlie control theoretical models are {\em independent of the specific input}, thus they can overcome the limitation of the control error to serve as oracle.
In combination to the falsification of the design assumptions, as mentioned in Section~\ref{sec:introduction}, we also want to avoid trivial failures and target instead \emph{subtle failures} (as they are called in the literature~\cite{zhang:2024,Hildebrandt:2020}).
For example, if a drone flies upside-down, the physics models used for the control design lose validity, causing the drone to crash.
However, the drone was not designed to fly upside-down; thus, despite showing a failure, this test case is not interesting.
In practice, trivial failure scenarios correspond to large deviations from the desired behaviour, which can be identified by large values of the control error.

Based on this observation, we formulate our {\bf testing objective} as the generation of arbitrary input traces that \emph{falsify the linearity design assumption while avoiding trivial failures}, where we identify the latter as test case executions yielding large control errors.
 \begin{figure}
    \centering
    \includegraphics{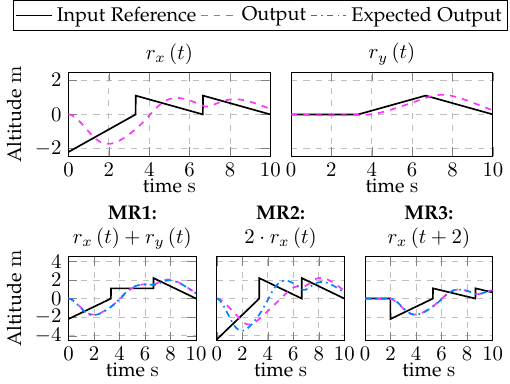}
    \caption{Example of MRs applications for a simplified drone altitude control example.
    Starting from the two initial tests in the first row ($r_x$ and $r_y$) we build three follow-up tests in the bottom row, showcasing MR1, MR2, and MR3, respectively.
    For the initial tests, we have only the input (black solid line) and output (purple dashed line) traces.
    Thanks to the MRs, for the follow up test cases we can also compute the expected output (dash-dotted lines) traces.
    }
    \label{fig:mrs-examples}
\end{figure}

\section{Metamorphic Relations}
\label{sec:metamorphic-relations}
Before delving into our approach, we introduce the MRs it relies upon.
As mentioned in Section~\ref{sec:introduction}, we define the MRs based on the definition of linear time-invariant (LTI) systems~\cite{Astrom:2008}.
A system $G$ is said to be LTI when it satisfies the three properties (PR) below for every input $r$.
In the properties definitions, $r_x\funof{t}$ and $r_y\funof{t}$ are input sequences functions of time $t$; $G[r\funof{t}]\funof{t}$, also a function of time $t$, is the system output in response to the input $r\funof{t}$; \asv\ and \tsv\ are real-valued constants.
\begin{enumerate}[label=\textbf{PR\arabic*:},itemindent=15pt]
    \item The output generated by an input that is the sum of two inputs, is equal to the sum of the outputs generated by the two inputs alone. Mathematically:
    \begin{equation*}
        G[r_x\funof{t}+r_y\funof{t}]\funof{t} = G[r_x\funof{t}]\funof{t} + G[r_y\funof{t}]\funof{t}.
    \end{equation*}
    \item The output generated by an input that is scaled by a constant, is equal to the output generated by the non-scaled input but scaled by the same quantity. Mathematically:
    \begin{equation*}
        G[\alpha \cdot r_x\funof{t}]\funof{t} = \alpha \cdot G[r_x\funof{t}]\funof{t}.
    \end{equation*}
    \item The output generated by an input that is shifted in time, is the same as the output generated by the non-shifted input but equally shifted in time. Mathematically:
    \begin{equation*}
        G[r_x\funof{t + \delta}]\funof{t} = G[r_x\funof{t}]\funof{t + \delta}.
    \end{equation*}
\end{enumerate}

These properties describe how the output of an LTI system is expected to change when applying specific change patterns to its inputs.
In the software testing literature, properties defined over sets of inputs and corresponding System-Under-Test (SUT) outputs are called MRs~\cite{chen:2018}.
Based on the three defining properties of an LTI system, we derive three MRs.
Our MRs are logical implications with an \emph{antecedent} that describes the input change pattern and a \emph{consequent} that describes how the output is expected to change.
\begin{enumerate}[label=\textbf{MR\arabic*:},itemindent=15pt]
    \item If an input is the sum of two other inputs, then its output is the sum of the two initial outputs:
    \begin{equation*}
        r\funof{t}=r_x\funof{t}+r_y\funof{t} \Rightarrow
    \end{equation*}\begin{equation*}
        G[r\funof{t}]\funof{t} = G[r_x\funof{t}]\funof{t} + G[r_y\funof{t}]\funof{t}.
    \end{equation*}
    \item If an input is equal to another input scaled by a factor \asv, then its output is the same as the initial output scaled by the same factor:
    \begin{equation*}
        r\funof{t} = \alpha \cdot r_x\funof{t} \Rightarrow G[r\funof{t}]\funof{t} = \alpha \cdot G[r_x\funof{t}].
    \end{equation*}
    \item If an input is equal to another input shifted in time by a step \tsv, then its output is the same as the initial output shifted in time by the same quantity:
    \begin{equation*}
        r\funof{t} = r_x\funof{t + \delta} \Rightarrow G[r\funof{t}]\funof{t} = G[r_x\funof{t}]\funof{t + \delta}.
    \end{equation*}
\end{enumerate}

The three MRs define equalities between input traces in the antecedent and output traces in the consequent.
We use the equality between input traces (i.e., the antecedent) to construct follow-up input traces.
Then, we run the initial and follow-up tests, obtain the tests outputs, and use the consequent to assess whether the MR holds or not, thus effectively serving as oracle.
In practice, we use the output of the initial test cases to obtain the {\em expected output} $e\funof{t}$ of the follow-up test case, i.e., the expected output trace.
Then, if the expected output is equal to the {\em actual output}, we consider the MR satisfied, otherwise we consider it falsified.

We showcase this use of the MRs in Figure~\ref{fig:mrs-examples}, where we run tests on a simplified model of a drone's altitude control.\footnote{
    The model is simplified in the sense that it only simulates the vertical movement of the drone.
}
However, we note that the MRs equally apply in the multi-dimensional case, e.g., if we were to consider all three dimensions x,y,z in which the drone can move.
In the figure, we use the two initial test cases shown in the upper plots, $r_x$ and $r_y$, to construct three follow-up test cases in the lower plots.
Each follow-up test case is obtained applying one of the MRs.
For the initial test cases we depict the input traces (i.e., the reference) as solid black lines and the output traces obtained after executing them as purple dashed lines.
For the follow-up plots, we also depict the expected output traces, built using the consequent of the MRs, with dash-dotted blue lines.
For MR1, we obtain a new input trace by summing (superimposing) the two initial input traces, shown in the bottom-left plot.
The expected output trace is also obtained by summing the outputs traces of the initial test cases.
Since the actual output matches the expected one, we conclude that MR1 is satisfied.
In contrast, in the bottom-center plot, we apply MR2, multiplying the input and output traces by 2.
In this case, we observe that the actual and expected output do not match, thus {\em falsifying} the MR.
In the bottom-right plot, we apply MR3 shifting in time the test case.
We thus deduce that MR3, like MR1, is satisfied.

As showcased in the examples, we assess the MRs' validity by comparing the expected and actual output traces of the follow-up test cases.
The higher the similarity between the two traces, the higher the degree of satisfaction of the MR.
In practice, we use a distance metric $d$ to define an \emph{MR-falsification degree} $\mrfalsify$:
\begin{equation}
    \mrfalsify=d\funof{G[r\funof{t}]\funof{t},e\funof{t}} .
    \label{eq:mr-falsify}
\end{equation}
For example, in MR2, the follow-up input reference is $r\funof{t}=\alpha\cdot r_x\funof{t}$, and its expected output is $e\funof{t} = \alpha \cdot G[r_x\funof{t}]\funof{t}$.
Then, to assess the validity of MR2, we compare $G[\alpha\cdot r_x\funof{t}]\funof{t}$ with $\alpha \cdot G[r_x\funof{t}]\funof{t}$.
The lower the distance between the traces, the lower the degree of MR-falsification, the more the MR is satisfied.
Ideally, when the MR is satisfied, we expect a zero distance, i.e., the actual and expected output traces are identical.
In practice, however, small deviations are expected because of phenomena such as measurement noise or uncertainty in the system.

Finally, we note that MRs can be applied in sequence; i.e., MRs can be applied to input and expected output traces that were also obtained from MRs.
In this way we can generate a wider variety of follow-up test cases.
For example, we can apply MR1 to two test cases where we first apply MR2 and MR3, respectively.
Such follow-up input trace $r_z\funof{t}$ looks as follows:
\begin{equation}
    r_z\funof{t} = \alpha \cdot r_x\funof{t} + r_y\funof{t+\delta} .
    \label{eq:mrs-example}
\end{equation}
Then, we can compare the output to assess whether the MRs are satisfied or not:
\begin{equation*}
    \mrfalsify=d(G[r_z\funof{t}]\funof{t}, \alpha \cdot G[r_x\funof{t}]\funof{t} + G[r_y\funof{t}]\funof{t+\delta}).
\end{equation*}
This subsequent application of MRs is commonly referred to as \emph{MR composition} in the MT literature~\cite{Qiu:2022}.
However, while MR-composition has been previously used to increase cost-effectiveness (i.e., composing the MRs to reduce the number of needed test cases while maintaining the MRs fault-finding capabilities), in our case, we use it instead to generate a wider variety of follow-up input traces.

When composing the MRs, we highlight that applying MRs in sequence does not require the actual output traces for the intermediate input traces to obtain the expected output.
Indeed, the expected output can be derived by applying the MRs in the sequence to the initial output traces, thus avoiding the need to execute the CPS multiple times.
In the example above in Equation~\ref{eq:mrs-example}, we limit the CPS executions to only three tests, to obtain $G[r_x\funof{t}]\funof{t}$, $G[r_y\funof{t}]\funof{t}$ and $G[r_z\funof{t}]\funof{t}$.
We do not need to run tests to obtain the intermediate results for $G[\alpha\cdot r_x\funof{t}]\funof{t}$ and $G[r_y\funof{t+\delta}]\funof{t}$, to separately assess the validity of the application of MR2 and MR3.
This allows us not to increase the number of required CPS executions when applying multiple MRs in sequence.
As a result, when using MRs composition, the number of test cases to be executed is equal to the number of initial test cases used, plus the one follow-up test case generated with the composed MRs; it is independent of the number of composed MRs.

\begin{figure}
    \centering
    \includegraphics{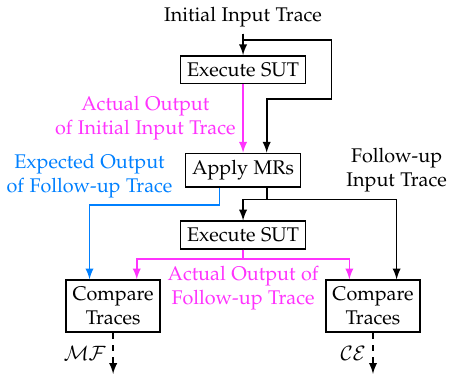}
    \caption{Use of the metamorphic relations in the proposed approach.
    Boxes represent the main steps of our approach; solid arrows represent the tests' input and output traces.
    Specifically, black arrows are the input traces, while the purple and azure arrows are the actual and expected output traces, respectively.
    The dashed arrows represent the distance metrics used to quantify the control error and the MR-falsification degree.
    }
    \label{fig:mrs-use}
\end{figure}

\section{Genetic Programming based on Metamorphic Relations}
\label{sec:approach}
We aim to generate input traces that falsify the MRs (i.e., test cases with high MR-falsification degrees), while avoiding trivial fails (i.e., test cases with large control error).
Specifically, we use the MRs to modify and combine initial input traces and build follow-up input traces together with their associated expected output traces.
Then, for the follow-up input traces, we can compute the control error ($\ctrlerr$) and MR-falsification degrees ($\mrfalsify$).

Figure~\ref{fig:mrs-use} shows how we use the MRs to both generate test cases and assess their control error and MR-falsification degrees.
In the figure, the colours of solid arrows distinguish the input traces (black arrows), the actual output traces of the tests (purple arrows), and the expected output traces of the new tests (azure arrows).
The blocks represent the different steps of our approach.
The ``Execute SUT'' blocks receive reference input traces and generate actual output traces.
The ``Apply MRs'' block uses the MRs to manipulate pairs of input and output traces and generate follow-up input traces together with the associated expected output traces.
In the right ``Compare Traces'' block, we compare the follow-up input and its actual output to compute the control error $\ctrlerr$.
In the left ``Compare Traces'' block, we compare the expected and actual outputs to assess whether the MRs hold or not for the given test cases (i.e., we compute the MR-falsification degree $\mrfalsify$, see Equation~\ref{eq:mr-falsify}).
The dashed arrows indicate the values obtained by the traces comparison.

\begin{algorithm}[tb]
    \caption{Evolutionary Search}
    \label{alg:search}
    \begin{algorithmic}[1]
        \Require pop\_size, num\_gens
        \Ensure archive
        \State pop$\gets$ generateInitialPopulation(pop\_size) \label{alg:search:gen-init}
        \State fitness$\gets$ assess(pop)                      \label{alg:search:ass-init}
        \State archive.update(pop,fitness)                     \label{alg:search:arc-init}
        \State i$\gets$0
        \While{i\textless num\_gens}                           \label{alg:search:loop}
            \State offspring$\gets$ breeding(pop)              \label{alg:search:breed}
            \State fitness$\gets$ assess(offspring)            \label{alg:search:ass}
            \State archive.update(offspring, fitness)          \label{alg:search:arc-upd}
            \State pop$\gets$ survival(offspring)              \label{alg:search:survive}
            \State i$\gets$ i+1
        \EndWhile
        \State\Return archive                                  \label{alg:search:return}
    \end{algorithmic}
\end{algorithm}

To generate the input traces, we search the space of possible MRs compositions applied to diverse initial input traces, i.e., the different ways to implement the ``Apply MRs'' block in Figure~\ref{fig:mrs-use}.
To perform this search, we leverage Genetic Programming (GP) to create programs that represent ways to compose and apply the MRs.
GP is an evolutionary approach that aims to search for the best program to perform a given task.
GP evolves a population of individuals where each individual is a program; the ability of a program to perform the given task is called fitness.
Like any evolutionary search algorithm, this population evolves over a number of generations through breeding (which entails crossover and mutation) and survival (selection of the fittest individuals).

Algorithm~\ref{alg:search} outlines the main steps of the evolutionary search algorithm that we employ.
The algorithm starts with the generation of the initial population (Line~\ref{alg:search:gen-init}) composed of a user-defined number of individuals (pop\_size).
Then, the fitness of the initial population is assessed (Line~\ref{alg:search:ass-init}) and the archive---which is the collection of the fittest individuals---is initialised with the fittest individuals of the initial population (Line~\ref{alg:search:arc-init}).
Afterwards, the main search loop (Line~\ref{alg:search:loop}) starts, which is executed as many times as the allotted generations (num\_gens).
For each generation, a new set of individuals is created by breeding the current generation (Line~\ref{alg:search:breed}).
This includes mutation and crossover, which are usually applied based on user-defined probabilities.
The offspring is then assessed (Line~\ref{alg:search:ass}) and used to update the archive (Line~\ref{alg:search:arc-upd}).
Finally, a subset of the offspring individuals is selected for survival and used to create the next generation (Line~\ref{alg:search:survive}).
After evolving the desired number of generations, the algorithm returns the archive with the fittest individuals to the user.

In our case, the objective is the maximisation of the MR-falsification degree while constraining the control error.
Thus, we define programs (i.e., individuals in search) that apply various combinations of MRs to different initial input traces and MRs parameters (i.e., the scaling value \asv\ when applying MR2 and the shift value \tsv\ when applying MR3).
We now first delve into the details of our GP-based test case generation, and then discuss how we apply combinations of MRs to generate arbitrary input traces together with the associated expected output traces.

\subsection{Genetic Program}
\subsubsection{Individual Representation and MR-Application Grammar}
\label{sec:individual-representation}
Our individuals are programs representing valid sequences of MRs applications (i.e., MRs compositions).
One of such programs starts from a number of initial input traces and returns a single input trace with an associated expected output trace.
To apply MR2, it needs scaling values for the variable \asv---how much to scale the trace---and, to apply MR3, it needs shift values \tsv---how much to shift the trace in time.
Accordingly, we identify three types of terminal symbols for the program grammar:
\begin{itemize}
    \item[\trc] is an initial input trace,\footnote{For brevity, from now on we drop the time-dependency $\funof{t}$ of the traces.}
    \item[\asv] is a scaling value used to apply MR2, and
    \item[\tsv] is a shift value used to apply MR3.
\end{itemize}
These terminal symbols are placeholders for actual input traces, scaling values, and shift values.
Technically speaking, the input traces and values are the actual terminal symbols.
For simplicity, we treat these placeholders as the terminal symbols and implement them as {\em ephemeral constants}.
Each instance of an ephemeral constant is assigned a new random value upon the program creation; this value is then kept for the rest of the search.
In this way, different instances of the same terminal have different values, even though from the point of view of the search they are still treated as constants.

Using these terminal symbols, we define the following syntax that represents the application of the MRs to arbitrary traces with arbitrary scaling or shifting values.
\begin{grammar}
<P> ::= <trace>

<trace> ::= \trc
 \alt  <trace>\ \SP\ <trace>
 \alt  \asv\ \AS\ <trace>
 \alt \tsv\ \TS\ <trace>
\end{grammar}

This grammar states that any valid program \synt{P} must return a trace.
A trace \synt{trace} can be a trace \trc\ by itself (potentially multi-dimensional), or it can be obtained from the application of one of the three MRs.
The first MR, denoted by \SP, generates a new trace by superimposing two traces \synt{trace}.
The second MR, denoted by \AS, generates a new trace from a scaling value \asv\ and a trace \synt{trace}.
The third MR, denoted by \TS, generates a new trace from a shift value \tsv\ and a trace \synt{trace}.
The follow-up trace is then computed by applying the associated MR.
As an example, the sequence of MR applications from Equation~\ref{eq:mrs-example} can be written as the following program \texttt{P}: 
\begin{center}
    \texttt{P=(\asv\AS$r_x$)\SP(\tsv\TS$r_y$)}.
\end{center}
The output of this program is then the follow-up trace $r_z$ of Equation~\ref{eq:mrs-example}.

\subsubsection{Generation of Initial Population}
We generate the initial population as a set of random programs with valid syntax.
To prevent the generation of trivial programs (e.g., an individual that is just a terminal symbol, like \texttt{P}$=r$, thus not representing the application of any MRs) or unduly large programs, we respectively set a minimum and maximum number of nested non-terminal symbols.
This ensures that the number of applied MRs in an individual is bounded by minimum and maximum limits.

\subsubsection{Fitness Assessment}
\label{sec:fitness-definition}
The objective of the search is the {\em maximisation of the MR-falsification degree while constraining the control error}.
Indeed, we are not interested in the minimization of the control error but rather in keeping it under an acceptable threshold.
Thus, we formulate the optimisation problem as a single objective (maximising the MR-falsification degree) with a constraint (keeping the control error below a threshold).
We use the penalty-method for constraint implementation~\cite{boyd:2004}, and define a fitness function that grows linearly with the MR-falsification degree but is exponentially penalised by the control error when it exceeds a given threshold.
For a given individual $i$, the fitness function is
\begin{equation}
    \fitness\funof{i} = \frac{\mrfalsify\funof{i}}{b^{c(\ctrlerr\funof{i}-\ctrlerrth)}},
    \label{eq:fitness}
\end{equation}
where $\mrfalsify\funof{i}$ is the individual's MR-falsification degree, $\ctrlerr\funof{i}$ is its control error, $\ctrlerrth$ is the threshold defining the maximum acceptable error, and $b$ and $c$ are user-defined positive constants.
The control error is the distance between the input reference and the actual output, thus
\begin{equation}
    \ctrlerr\funof{i} = d\funof{r,G[r]}.
    \label{eq:ctrl-err-def}
\end{equation}
The MR-falsification degree is computed based on Equation~\ref{eq:mr-falsify}, where we obtain the expected output based on the actual output of the terminal symbols traces \trc.
As mentioned at the end of Section~\ref{sec:metamorphic-relations}, we do not need to compute the output for the intermediate test cases.

When computing $\mrfalsify$ and $\ctrlerr$, we want to assess how close the values of the expected and actual output traces are, at each point in time, and similarly the reference input and actual output traces, respectively.
To quantify this distance, in our approach, we use the Euclidean distance (commonly used in previous CPS literature to compare traces~\cite{Matinnejad:2019}), and divide it by the number of dimensions (in case of multi-dimensional traces) and the length of the trace.\footnote{
    Since all the traces are of the same length and dimension, the division is not strictly necessary.
    However, it makes the metric equal to the average distance, thus enabling a more intuitive interpretation of the metric.
}
Mathematically, for two multi-dimensional traces $a$ and $b$, with values $a_{i,k}$ and $b_{i,k}$ for dimension $i$ at time step $k$, we compute
\begin{equation}
    d\funof{a,b}=\frac{\sum_{k}\sqrt{\sum_{i}\funof{a_{i,k}-b_{i,k}}^2}}{n_{\mathit{dim}}k_{\mathit{max}}},
    \label{eq:euclidean-distance}
\end{equation}
where $n_{\mathit{dim}}$ is the number of dimensions, and $k_{\mathit{max}}$ is the number of samples of the traces.
Using such a metric, we can then use a threshold to assess whether two traces are similar or not.
This distance can be interpreted as the average value difference between the two traces over time steps.
For example, two traces with a distance of 5 means that, at each time step, they differ on average by 5 units.

\subsubsection{Breeding: Mutation and Crossover}
\label{sec:mutation-and-crossover}
For the mutation and crossover operators, we employ standard solutions from the GP literature.
When mutating an individual (program), we randomly select a terminal or non-terminal symbol.
We then substitute it (and the symbols that it expands into) with a new randomly generated sub-program that returns a compatible symbol with the one removed.
For example, in the program shown above, we could select \AS\ and substitute it with the \SP\ symbol and new randomly generated ephemeral constants.
The program \texttt{P} is then mutated into \texttt{P$_\text{mut}$}, where the azure (underlined) part is mutated into the purple (underlined) one:
\begin{center}
    \texttt{P=\textcolor{exp}{\underline{(\asv\AS$r_x$)}}\SP(\tsv\TS$r_y$)\\
    $\rightarrow$\\
    P$_\text{mut}=$\textcolor{act}{\underline{($r_w$\SP$r_j$)}}\SP(\tsv\TS$r_y$)}.
\end{center}

Similarly, when performing a crossover, we randomly select a terminal or non-terminal symbol from one of the two individuals selected for reproduction.
We then select (again randomly) a compatible symbol from the other individual, where compatible means that the two symbols can be swapped while still obtaining syntactically valid programs.
The crossover is then performed by swapping the two symbols in the two individuals.
An example of crossover is
\begin{center}
    \texttt{P$_\text{1}$=\textcolor{exp}{\underline{(\asv$_1$\AS$r_x$)}}\SP(\tsv\TS$r_y$)}, 
    \texttt{P$_\text{2}$=\asv$_2$\AS(\textcolor{exp}{\underline{$r_w$}}\SP$r_j$)}\\
    $\rightarrow$\\
    \texttt{P$_\text{3}$=\textcolor{act}{\underline{$r_w$}}\SP(\tsv\TS$r_y$)},
    \texttt{P$_\text{4}$=\asv$_2$\AS(\textcolor{act}{\underline{(\asv$_1$\AS$r_x$)}}\SP$r_j$)}.\\
\end{center}
Here, the parents \texttt{P$_\text{1}$} and \texttt{P$_\text{2}$} generate the offspring \texttt{P$_\text{3}$} and \texttt{P$_\text{4}$} by swapping (\asv$_1$\AS$r_x$) with $r_w$: the symbols involved in the swap are underlined, and highlighted in azure for the parents and in purple for the offspring.

\subsubsection{Selection Mechanism and Diversity}
\label{sec:selection-mec-and-diversity}
According to the GP algorithm, at each generation, we mutate and crossover the current population to create an offspring.
Among the children in the offspring, we then select the fitter ones to be carried over to the next generation.
For this selection, we use the standard selection tournament, where a given number of individuals (usually two or three) are randomly picked, and among those the one with the highest fitness is passed on to the next generation.
This is repeated until enough individuals have been selected to form the new generation.

To avoid duplicates, before adding an individual to the new generation, we compare it with individuals that are already included in the next generation.
We compare two individuals by measuring the Euclidean distance between the input traces that they generate, as for $\mrfalsify$ and $\ctrlerr$ in Equation~\ref{eq:euclidean-distance}.
If the two programs generate input traces with a distance lower than a user-defined {\em similarity threshold}, we consider the two individuals to be the same, and we do not include the new one in the new generation.
When an individual is considered a duplicate, we discard it and add instead a random individual from the archive.
In this way, we avoid having similar individuals in each generation, thus maintaining population diversity.

\subsubsection{Archive and Search Output}
\label{sec:archive}
At the end of the search, we return to the user an {\em archive} containing high-fitness individuals generated through the whole search.
To avoid duplicates in the archive, we add a new individual to the archive only if it is sufficiently different from the ones already contained in it.

Like for diversity in the selection mechanism, we use the Euclidean distance to compare individuals, and the same user-defined threshold to assess if two individuals are too similar and can be considered to be the same.
When an individual is found to be similar to another already in the archive, even if the new individual has higher fitness, this implementation will keep the one already present in the archive.
Even if this can result in keeping the one with lower fitness among the two, the two individuals are similar enough for this not to matter and \emph{they will not show different behaviours of the CPS}.
Thus, in practice, the choice of keeping in the archive the individual already in it does not affect the overall output of the search.

\subsection{Generation of Valid Arbitrary Input Traces}
\label{sec:gen-valid-traces}
We conclude the description of our approach by discussing how we generate values for the terminal symbols and apply combinations of MRs in practice.
This is important as it affects both the effectiveness of the search as well as which traces can be generated with our approach.

First, we want the initial input traces of a program to be within the design scope (which, as mentioned in Section~\ref{sec:introduction}, is the set scenarios where the linear model design assumption is valid).
In this way, the expected output trace of the program is representative of the CPS behaviour as predicted by the control theoretical models.
Thus, the MR-falsification degree is then a measure of how far the follow-up input trace is from the design scope and can be used to effectively guide the search.
Importantly, to obtain expected outputs that are representative of the CPS behaviour within the design scope, we also rely on the choice to not execute the intermediate test cases.
As mentioned in the fitness definition~\ref{sec:fitness-definition}, we compute the expected outputs based only on the output of the terminal symbols traces, which always have small amplitude and simple shapes.
Then, since the test cases associated with the terminal symbols traces can be assumed to be within the design scope, the expected output of the follow-up test case will be representative of the CPS behaviour within the design scope.

Second, we want our approach to be able to generate varied and arbitrary input traces within a user-defined valid input range.
The use of this valid input range allows users to avoid input values that are not relevant to the CPS under test.
Within this range of valid inputs, we then want to enable the generation of arbitrary input traces, which is important for the search to comprehensively explore the input space.
This requires covering the whole range of input amplitudes combined with the generation of diverse patterns.

\begin{figure}
    \centering
    \includegraphics{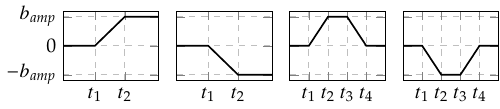}
    \caption{The four different patterns used for the definition of the initial traces. Each pattern has a number of time parameters (the $t_i$ indices) that are randomly selected when the ephemeral constant is created for a initial trace terminal symbol.
    In contrast, $b_{\mathit{amp}}$ is fixed and set by the user to ensure that the initial trace belongs to the design scope.}
    \label{fig:base-traces-patterns}
\end{figure}

\subsubsection{Generation of the Initial Input Traces}
\label{sec:initial-traces-generation}
We ensure that the initial input traces (i.e., the ephemeral constants of the terminal symbol \trc) are within the design scope by using simple input patterns and small amplitude values.
As discussed in our previous work~\cite{mandrioli:2023b}, simple patterns and especially small amplitudes are not expected to falsify the linearity assumptions.
Intuitively, in a CPS, a short and slow movement is simpler to handle than a long and fast one.

We use the patterns shown in Figure~\ref{fig:base-traces-patterns} with a fixed user-defined amplitude $b_{\mathit{amp}}$.
To guarantee that the initial inputs are within the design scope, $b_{\mathit{amp}}$ can be appropriately chosen.
The patterns include linear increments and decrements of the values.
The idea is that by applying a combination of the MRs to an adequate number of such initial traces, we can, in theory, approximate any arbitrary trace.
If the CPS does not perform well with these proposed patterns, it indicates that there are some fundamental issues with the CPS design.
This further suggests that it is too early in the development to start testing with more complex traces.
For example, if a drone is not able to move for one meter in a straight line, it is too early to test it with more complex or longer sequences.
In these simple patterns, the parameters $t_1$, $t_2$, $t_3$ and $t_4$ are randomly selected each time an ephemeral constant is generated for each initial input trace symbol \trc.
By allowing different values for such parameters, we enable the generation of diverse initial input traces that can then be combined with the MRs and generate more complex follow-up traces.

As visible from Figure~\ref{fig:base-traces-patterns}, the initial input traces are centred around zero.
In practice, before feeding the input traces to the SUT, we apply an {\em input bias} that shifts them in the middle of the user-defined valid input range.
In this way, the patterns can be seen as deviations from the input bias rather than absolute values.
This allows us to guarantee that all initial input traces are valid (as long as $b_{\mathit{amp}}$ is within the valid range) and to treat cases where the valid input range is not centred around zero in the same way as cases where it is.

\subsubsection{Implementation of MR2 to Cover the Valid Input Range}
To generate input traces within the valid input range, we use the amplitude scaling of MR2, since the initial input traces have a fixed amplitude.
However, at the same time, we need to ensure that we do not exceed the valid input range.
This implies that the maximum allowed value for the scaling depends on the amplitude of the trace that it is applied to.
For example, if the user-defined valid range is between 2 and $-2$, and the maximum and minimum values of the trace that we are applying the MR to are 1 and $-0.5$, then the set of allowed scaling values that will not generate an invalid trace is between 0 and 2.
However, if the maximum value of the trace was 2, then the maximum scaling value would be 1.
To make sure that we can both cover the valid input range but also that we do not exceed it, we use for the symbol \asv\ a value scaled between 0 and 1.
Then, when we apply MR2, we first compute the set of valid scaling values for the given trace and then use the \asv\ value to select the actual scaling value from this set.
In the example above, with valid scaling values between 0 and 2, an \asv\ value of 0.5 would lead to the selection of the actual scaling value of 1, and 0.33 would lead to the selection of 0.66.

For some corner cases, when the trace to which the MR is applied has only small values, the set of valid amplitude scaling values can include some very large ones.
For example, in a drone, if the maximum value of the trace is \qty{0.001}{\metre} and the maximum valid amplitude value is 2, the maximum allowed scaling value is 2000.
While this is correct and will generate valid inputs, it can cause problems when computing the expected output.
In fact, if the output of the initial trace has even a small deviation (e.g., because of some noise in the sensors), when scaled by a large value, this deviation will be amplified.
For example, a deviation by \qty{0.01}{\metre}, that can be expected because of noise, might be scaled by 2000, and generate an expected output that is in the range of \qty{20}{\metre}, while the reference is still within \qty{2}{\metre} and \qty{-2}{\metre}.
This is technically correct and can help detecting unexpected CPSs behaviours\footnote{
    In the example, it would highlight that the \qty{0.01}{\metre} initial deviation was not expected.
} but, due to the large deviation of the expected output, it can generate individuals with such a high fitness that the search will focus solely on them.
Indeed, given the large deviation of the expected output from the reference, the fitness of these individuals can be orders of magnitude larger than the other individuals.

To avoid such extreme cases, we limit the actual scaling values to the ratio between the maximum allowed amplitude and the amplitude of initial traces.
In the drone example with maximum amplitude of \qty{2}{\metre}, if $b_{\mathit{amp}}$ was \qty{0.02}{\metre}, then we would limit every scaling value to 100.
With this rule of thumb, we can generate all the inputs within the allowed range but avoid extreme amplitude scaling values.

\subsubsection{Implementation of MR1 and MR3 to Generate Diverse Patterns}
To obtain input traces with more varied and arbitrary patterns, we leverage MR1 and MR3 to combine the initial input traces.

MR1 combines different traces by summing them, and thus can effectively create new arbitrary patterns.
However, when applying MR1 to two traces, the amplitude values of the new trace will likely increase and might exceed the valid range.
To make sure MR1 does not generate inputs that are out of the valid range, we always apply it in combination with MR2 and divide by 2 the new trace.
In this way, the new trace is always within the same range of values as the original traces and cannot violate the user-defined range.

As for MR3, while by itself it does not alter the pattern of a given input to which it is applied (the trace is only shifted in time, and the values do not change), it enables the combination of two traces, through MR1, in different ways by using different shift values.

This synergy between MR3 and MR1 provides an extra degree of freedom to the search algorithm when combining two traces.
As a result, we can rely less on the variety of initial input traces and leverage more the search algorithm for exploring the space of possible input traces.
 \section{Empirical Evaluation}
\label{sec:empirical-evaluation}
Our approach aims to generate input traces, together with the associated oracles (in the form of the expected output), that falsify the MRs.
Accordingly, in our empirical evaluation, we assess our contribution both to the generation of such input traces and to the oracle problem.
First, we assess whether the GP search is actually needed as well as its effectiveness at generating non-trivial traces that falsify the MRs.
Second, we assess the ability of the MRs to extend the state of the art (i.e., methods based on the control error) as oracle for CPSs.
We formulate the objectives of our experiments with the following research questions (RQs):
\begin{itemize}[itemindent=15pt]
    \item[\textbf{RQ1:}] How does the GP search compare to a baseline based on random generation of programs?
\end{itemize}
With this RQ, we assess the GP search part of the proposed test case generation approach, i.e., its ability to generate input traces that falsify the MRs but are not trivial failures.
We compare the achieved fitness against the random generation of programs in order to assess the need for an evolutionary algorithm.
In this way, we can assess how difficult it is to falsify the MRs while avoiding very large control error values (i.e., trivial failures).

\begin{itemize}[itemindent=15pt]
    \item[\textbf{RQ2:}] How can MR-falsification improve oracles that rely solely on control error?
\end{itemize}
With this RQ, we evaluate whether the MR-falsification metric provides engineers with new information when compared to the sole use of the control error.
Specifically, we are interested in assessing whether, for similar control error values, we obtain different MR-falsification values.
If so, the MR-falsification metric offers engineers a new mechanism to distinguish between passed and failed test cases.

Before addressing our RQs, we present the test subjects and the settings chosen for the GP.

\subsection{Test Subjects}
\label{sec:test-subjects}
To evaluate our approach, we need test subjects that are developed with the use of linear systems theory, and include both the software and a simulator of the physical part.
In this way, we can test the interaction with the physics.
Under these requirements, we identified three test subjects, further described below: the Crazyflie drone, the Engine Timing Model with Closed-Loop Control, and the Lightweight Aircraft Design.

The Crazyflie drone (CF) is a drone used both in robotics~\cite{greiff:2021} as well as in software testing research~\cite{mandrioli:2023a,mandrioli:2023b}.
We rely on a CF simulation model from previous research, which consists of a simulator of the physical drone flying in space, connected in closed-loop with a Python implementation of the control software.
It receives as input the desired position expressed as the three coordinates x, y, z, and it simulates how the control layer interacts with the physics to bring the drone to the desired position.
The CF control algorithm is composed of an Extended Kalman Filter and a total of eleven PIDs~\cite{Astrom:2008}.
The filter uses sensor data to estimate the position, speed, attitude and attitude rate of the drone, which the PIDs use in turn to compute the actuation commands sent to the motors.\footnote{
    More details on the CF control algorithm can be found at \url{https://www.bitcraze.io/documentation/repository/crazyflie-firmware/master/functional-areas/sensor-to-control/controllers/}.
}

The Engine Timing Model with Closed-Loop Control (ET) controls the revolutions per minute (rpm) of an internal combustion engine.
Here, the software controls the fuel injection to bring the engine speed to the desired one.
The simulation model used for the testing is developed by Mathworks and is implemented in Simulink, a common tool for the development of CPSs.
The ET control algorithm is a PID which is executed based on the crankshaft and which, based on the desired and actual rpm of the motor, controls the opening of the throttle.\footnote{
    More details on the test subject and its control algorithm can be found at \url{https://www.mathworks.com/help/simulink/slref/engine-timing-model-with-closed-loop-control.html}.
}

The Lightweight Aircraft Design (LA) comprises two controllers dedicated to controlling the pitch of the aircraft (also known as the angle of attack) and to controlling the aircraft altitude, respectively.
The two controllers are implemented in a so-called cascaded control architecture~\cite{Astrom:2008}, where the altitude controller uses the pitch controller (intuitively, changing the pitch the aircraft makes the aircraft gain or lose altitude).
Both controllers are defined as transfer functions (a generalisation of PIDs).
A pressure sensor is used to measure the aircraft current altitude and an inertial measurement unit (accelerometer and gyroscope) is used to estimate the pitch angle.\footnote{
    More details on the test subject and its control algorithm can be found at \url{mathworks.com/help/aeroblks/lightweight-airplane-design.html}.
}

\subsection{Parameter Settings}
\label{sec:parameter-setting}
The application of our testing approach requires setting parameters for
\begin{enumerate*}[label=(\roman*)]
    \item the implementation of the MRs, and
    \item the GP algorithm.
\end{enumerate*}

\subsubsection{Setting MR Implementation Parameters with Guidelines}
\begin{table}[t]
    \centering
    \caption{Parameters required for the implementation of the MRs for each test subject based on our engineering guidelines.}
    \begin{tabular}
    {p{0.285\columnwidth}
     p{0.23\columnwidth}
     p{0.15\columnwidth}
     p{0.15\columnwidth}}
     \toprule
    {\bf Parameters}  & {\bf Values CF}   & {\bf Values ET}    & {\bf Values LA}    \\\midrule
    Tests duration    & \qty{10}{\second} & \qty{50}{\second}  & \qty{120}{\second} \\
    Warm-up time      &  \qty{3}{\second} & \qty{1.5}{\second} & \qty{1}{\second}   \\
    Initial amplitude & (\qty{0.2}{\metre},\qty{0.2}{\metre},\qty{0.2}{\metre}) & \qty{500}{\rpm} & \qty{50}{\metre} \\
    Amplitude range   & $x,y$:(\qty{2}{\metre},\qty{-2}{\metre}) $z$:(\qty{0.5}{\metre},\qty{1.2}{\metre}) & (\qty{6000}{\rpm}, \qty{1200}{\rpm}) & (\qty{2300}{\metre}, \qty{1700}{\metre}) \\
    \bottomrule
\end{tabular}
     \label{tab:parameters-mrs-set}
\end{table}

Table~\ref{tab:parameters-mrs-set} lists the parameters that need to be set for the MRs implementation together with the values chosen for our test subjects.
These parameters have a quantifiable impact on the approach implementation for a given test subject; thus we can provide and use a rationale for their setting.
We provide the following guidelines to set them:
\begin{itemize}
    \item {\bf Test duration}: Selecting the test duration involves a trade-off between testing effectiveness and cost.
    The tests should be as short as possible to reduce testing cost, but long enough to allow the CPS to perform manoeuvres of interest.
    For example, for the delivery drone, the duration should be comparable to the time required to fly between a couple of way-points of the planned path.
    \item {\bf Warm-up time}: This parameter sets a warm-up time for the CPS to reach a state where the actual test can start (e.g., a drone has to take off).
    In practice, it sets an initial time in which the input is fixed to the input bias (defined in Section~\ref{sec:gen-valid-traces}) so that the CPS can reach that state.
    It can be set by analysing a test that inputs only the bias, observing how long it takes for the CPS to get ready for testing.
    At the same time, it can be kept as small as possible to minimise testing time.
    For the drone, this would be the time taken by the drone to take off and stably reach the desired altitude at the beginning of each test.
    \item {\bf Initial amplitude}: This parameter sets the amplitude values for the base test cases.
    As discussed in Section~\ref{sec:initial-traces-generation}, it should be small enough such that the CPS is expected to process it with the expected performance (e.g., speed), that is be within the design scope.
    At the same time, it should be large enough to actually perform a manoeuvre.
    In our delivery drone example, it should be large enough for the drone to actually move but small enough such that the initial test is within the design scope.
    \item {\bf Amplitude range}: This parameter sets the valid value range for each input trace.
    It can be set according to the maximum values that the CPS is required to handle.
    In the drone, this could be the size of the space that the drone is supposed to fly in.
\end{itemize}

\subsubsection{Search Parameters Settings}
\label{sec:search-parameters-setting}
For our search problem we use the $\funof{\mu,\lambda}$ algorithm, which is a standard evolution strategy in the meta-heuristic literature.
Setting the parameters for a search algorithm is a problem extensively studied in the literature.
It has been observed that there is no unique answer to such problem, which largely depends on the specific application~\cite{Arcuri:2011}.
For this reason, based on existing guidelines~\cite{Arcuri:2011}, we use manual tuning to find a working set of values for the GP search.

Table~\ref{tab:parameters-gp} lists the parameters that we set in this way, together with the corresponding values.
The table also shows the maximum number nodes that we allowed in any individual for bloating prevention (i.e., the prevention of the generation of individuals so large that they could not be handled by the machine running the experiments).
In our experiments, however, this number of nodes in an individual (300) was never reached, thus making unnecessary the bloating prevention mechanism.
Note that, for the parameters in Table~\ref{tab:parameters-gp}, we used the same values across all test subjects.

\begin{table}[t]
    \centering
    \caption{Parameters of the GP set with guidelines and manual tuning. We use the same values for each test subject.}
    \begin{tabular}
    {p{0.72\columnwidth}
     p{0.2\columnwidth}
    }
    \toprule
    {\bf Parameter} & {\bf Value} \\
    \midrule
    Algorithm & $\funof{\mi,\lambda}$ \\
    $\mi$ (individuals per generation)  & 50 \\
    $\lambda$ (offspring size at each generation) & 80 \\
    initial population size & $\mi$ \\
    crossover rate & 0.35 \\
    mutation rate & 0.35 \\
    tournament size & 2 \\
    number of generations & 40 \\
    max number of nodes for bloating prevention & 300 \\
    max initial individuals depth & 8 \\
    min initial individuals depth & 4 \\
    min and max depth of mutated subtrees & (2,4) \\
    \bottomrule
\end{tabular}
     \label{tab:parameters-gp}
\end{table}

The remaining parameters are specific to our search problem and need to be set for each application.
Such parameters include the coefficients of the fitness function (the exponential base $b$, the exponent scaling coefficient $c$, and the control error threshold $\ctrlerrth$), and the similarity threshold (Section~\ref{sec:selection-mec-and-diversity}) used to compare individuals.

The control error threshold defines the control error value above which we consider a test to be a trivial failure (with fitness being exponentially penalised above this value).
It represents the maximum accepted average distance between the actual output and the reference input.
At the limits, a very small value means considering every deviation from the input reference a trivial failure.
In this case, the search would focus on very simple tests where the CPS is able to closely track the input reference.
Conversely, a very large control error threshold would allow for large deviations from the reference inputs, and the maximisation of the MR-falsification degree would then favour the generation of trivial failures.
Different values in-between should lead to a trade-off between these behaviours; however, the specific values that can achieve this trade-off are dependant on the application.
Since we expect CF to reach a position with an accuracy in the order of centimetres, we consider average distances above \qty{0.15}{\metre} from the reference position to be trivial failures.
For ET, we expect it to track the reference speed with an accuracy around tens of rpm and we thus consider average distances above \qty{75}{rpm} from the reference rpm value to be trivial failures.
In the case of LA, the expected precision is in the order of meters;  we thus set the threshold at \qty{2}{\metre}.

The similarity threshold defines instead the minimum Euclidean distance below which two input traces are considered too similar and thus redundant.
Similarly to the control error threshold, this parameter also depends on the expected accuracy of the CPS in tracking an input.
Very high values lead to considering all tests similar to each other, thus leading to the generation of new random individuals to increase diversity.
At the limit, this leads to a behaviour equivalent to random search.
Conversely, for very low values, all the individuals are considered different from each other and the mechanism for increasing diversity is never triggered.
This would lead to the generation of very similar individuals, thus limiting the exploration of different CPS behaviours.
Again, different values in-between should lead to a trade-off between these behaviours; however, the specific values that achieve this trade-off depend on the given application.
Intuitively, the more accurate the CPS is expected to be, the more small input differences affect it.
Thus, for the similarity threshold, we use larger values than the control error threshold to make sure that the difference between two traces is relevant when considering the expected precision.
Specifically, we use \qty{0.2}{\metre} for CF, \qty{300}{rpm} for ET, and \qty{10}{\metre} for LA.
Table~\ref{tab:parameters-thresholds} summarises the threshold values chosen for our test subjects.

\begin{table}[t]
    \centering
    \caption{Control error and similarity thresholds chosen for each test subject.}
    \begin{tabular}
    {p{0.4\columnwidth}
     p{0.135\columnwidth}
     p{0.135\columnwidth}
     p{0.135\columnwidth}}
     \toprule
    {\bf Parameter}         & {\bf CF} & {\bf ET} & {\bf LA}\\
    \midrule
    Control error threshold & \qty{0.15}{\metre} & \qty{75}{\rpm}  & \qty{2}{\metre} \\
    Similarity threshold    & \qty{0.2}{\metre}  & \qty{300}{\rpm} & \qty{10}{\metre} \\
    \bottomrule
\end{tabular}
     \label{tab:parameters-thresholds}
\end{table}

The remaining coefficients to be tuned are the two coefficients of the fitness function: the base $b$ of the exponential and the scaling factor $c$ in the exponent of Equation~\ref{eq:fitness}.
These parameters implement the exponential penalisation of the control error with respect to the MR-falsification.
However, it is difficult to quantify a priori their impact on the search.
For this reason, we empirically set them according to three steps:
\begin{enumerate*}[label=(\roman*)]
    \item we define different sets of candidate values,
    \item we run experiments with the chosen sets of values, and
    \item we compare the individuals in the archive in terms of average control error and MR-falsification to select the set of parameters that achieve the best MR-falsification while maintaining the control error below a specified threshold $\ctrlerrth$.\end{enumerate*}

For the base $b$, we include two alternatives frequently used in the literature, $10$ and  Euler's number $\mathrm{e}$, as well as a smaller value of $1.5$.
The coefficient $c$ is a scaling factor for the control error contribution.
Since such a contribution depends on the unit of measure, it can have very different values in different applications---in our case, we have \qty{0.15}{\metre} in CF, \qty{75}{rpm} in ET, and \qty{2}{\metre} in LA.
Thus, we adjust it to the control error range and avoid making the exponential function always very large or very small.
Specifically, we use three values of $c$ so that $c\cdot\ctrlerrth$ equals 0.5, 1, and 5.

\begin{table}[t]
    \centering
    \caption{Results of the empirical investigation of the fitness function coefficients.
    The results include the average MR-falsification and control error of the archive.
    For our evaluation we used the parameters, highlighted in bold and purple, that achieved the highest $\mrfalsify$ while maintaining the $\ctrlerr$ below or close to $\ctrlerrth$.}
    \def\lClmn {8mm}
\newcommand{\w}[1]{\textcolor{act}{\bf #1}} \begin{tabular}{p{0.5\lClmn}p{\lClmn}p{\lClmn}p{\lClmn}p{\lClmn}p{\lClmn}p{\lClmn}p{\lClmn}p{\lClmn}p{\lClmn}}
\toprule
  \multicolumn{2}{c}{\multirow{2}*{\bf CF}}  & \multicolumn{6}{c}{ $b$ }                                     \\ \cmidrule{3-8}
                      &       & \multicolumn{2}{c}{1.5} &  \multicolumn{2}{c}{$\mathrm{e}$} & \multicolumn{2}{c}{10}  \\  \cmidrule(lr){3-4}  \cmidrule(lr){5-6} \cmidrule(lr){7-8}
                      &       & $\ctrlerr$ &$\mrfalsify$& $\ctrlerr$ &$\mrfalsify$ & $\ctrlerr$ &$\mrfalsify$\\ \cmidrule(lr){3-8}
    \multirow{3}*{$c$}&  3.33 &   0.681    &   0.677    &   0.283    &    0.311    &  0.136     &   0.131    \\ \cmidrule(lr){3-8}
                      &  6.66 &   0.324    &   0.335    &\w{0.146}   & \w{0.160}   &  0.084     &   0.073    \\ \cmidrule(lr){3-8}
                      & 33.33 &   0.112    &   0.124    &   0.053    &    0.033    &  0.049     &   0.023    \\ 
\bottomrule
\end{tabular}

\vspace{2mm}

\begin{tabular}{p{0.5\lClmn}p{\lClmn}p{\lClmn}p{\lClmn}p{\lClmn}p{\lClmn}p{\lClmn}p{\lClmn}p{\lClmn}p{\lClmn}}
\toprule
    \multicolumn{2}{c}{\multirow{2}*{\bf ET}} & \multicolumn{6}{c}{ $b$ }                                    \\ \cmidrule{3-8}
                      &       & \multicolumn{2}{c}{1.5} &  \multicolumn{2}{c}{$\mathrm{e}$} &  \multicolumn{2}{c}{10} \\ \cmidrule(lr){3-4}  \cmidrule(lr){5-6} \cmidrule(lr){7-8}
                      &       & $\ctrlerr$ &$\mrfalsify$& $\ctrlerr$ &$\mrfalsify$ & $\ctrlerr$ &$\mrfalsify$\\ \cmidrule(lr){3-8}
    \multirow{3}*{$c$}& 0.007 &   77.28    &   22.36    &   166.01   &  111.62     &   85.50    &   43.75    \\ \cmidrule(lr){3-8}
                      & 0.013 &  126.41    &   65.52    &    94.46   &   48.56     &   48.73    &   23.13    \\ \cmidrule(lr){3-8}
                      & 0.066 &\w{56.03}   & \w{24.73}  &    31.76   &    8.86     &   21.91    &    3.63    \\
\bottomrule
\end{tabular}

\vspace{2mm}

\begin{tabular}{p{0.5\lClmn}p{\lClmn}p{\lClmn}p{\lClmn}p{\lClmn}p{\lClmn}p{\lClmn}p{\lClmn}p{\lClmn}p{\lClmn}}
\toprule
    \multicolumn{2}{c}{\multirow{2}*{\bf LA}} & \multicolumn{6}{c}{ $b$ }                                    \\ \cmidrule{3-8}
                      &       & \multicolumn{2}{c}{1.5} &  \multicolumn{2}{c}{$\mathrm{e}$} &  \multicolumn{2}{c}{10} \\ \cmidrule(lr){3-4}  \cmidrule(lr){5-6} \cmidrule(lr){7-8}
                      &       & $\ctrlerr$ &$\mrfalsify$& $\ctrlerr$ &$\mrfalsify$ & $\ctrlerr$ &$\mrfalsify$\\ \cmidrule(lr){3-8}
    \multirow{3}*{$c$}&  0.25 &    10.684  &  9.632     &   5.925    &   4.736     &   1.382    &   0.758    \\ \cmidrule(lr){3-8}
                      &  0.50 &     5.946  &  4.981     &\w{2.103}   &\w{1.245}    &   0.909    &   0.450    \\ \cmidrule(lr){3-8}
                      &  2.50 &     1.007  &  0.671     &   0.589    &   0.350     &   0.560    &   0.172    \\
\bottomrule
\end{tabular}
     \label{tab:parameters-mrs-investigation}
\end{table}

Table~\ref{tab:parameters-mrs-investigation} shows the results of the tests runs for each test subject: each row corresponds to a value of $c$ and each column to a value of $b$.
In the table, we highlight in bold and purple the results that achieve the best $\mrfalsify$ while maintaining $\ctrlerr$ below the threshold, and therefore achieve the best falsification of the design assumptions while avoiding trivial failures.
For LA, since we did not obtain any value close to and lower than $\ctrlerrth$, we selected the parameters for which we get the closest value to the threshold: $c=0.5$ and $b=\mathrm{e}$.
For answering our RQs, we rely on those parameter values and use the associated testing results.

\subsubsection{Assessment of Search Random Variability}
\label{sec:approach-variablity-assessment}
Since the results can be affected by the random nature of meta-heuristic search, we repeated the experiments with the chosen parameters to assess the impact of randomisation on the algorithm.
With a time budget of around one day for each test subject, we repeated the execution of our approach ten times (on top of the executions used for the parameters assessment) with different seeds for the random number generator.
We then assessed if the average control error and MR-falsification degree of the individuals in the archive presented significant variance.

Across the ten runs, we observed: for CF, $\ctrlerr = 0.156\pm0.008$ and $\mrfalsify = 0.169\pm0.035$; for ET, $\ctrlerr = 54.36\pm4.5$ and $\mrfalsify = 24.9\pm3.97$; for LA, $\ctrlerr = 2.286\pm0.56$ and $\mrfalsify = 1.58\pm0.44$. 

These results correspond to a small variation in the metrics' values, in the order of millimetres and centimetres for CF, units of \unit{\rpm} for ET, and lower than a metre for LA.
This shows the limited impact of randomness on our approach execution.

To answer our research questions, we used the experimental results highlighted in bold and purple in Table~\ref{tab:parameters-mrs-investigation}.
We note that the repeated experiments, besides showing low random variability of the approach, also suggest that the selected runs (the ones from Table~\ref{tab:parameters-mrs-investigation}) are close to the average behaviour of the search, and they are no outliers.
We also remark that further tuning of the aforementioned parameters might improve the performance of our approach.
However, our current settings have already enabled us to convincingly and clearly support our conclusions.
Therefore, we do not report on additional experiments to optimise these values.

\begin{figure*}
    \centering
    \includegraphics{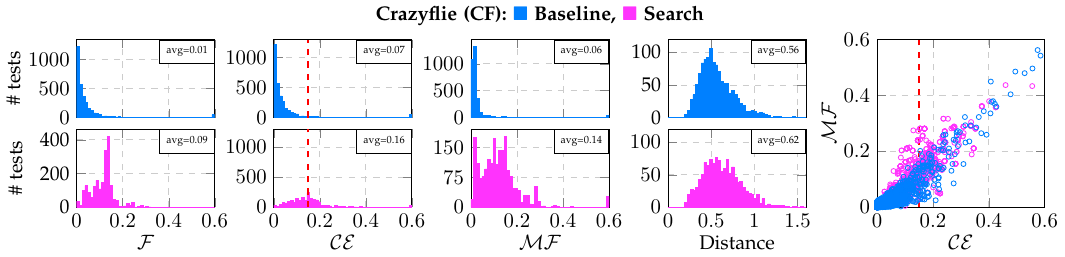}
    \includegraphics{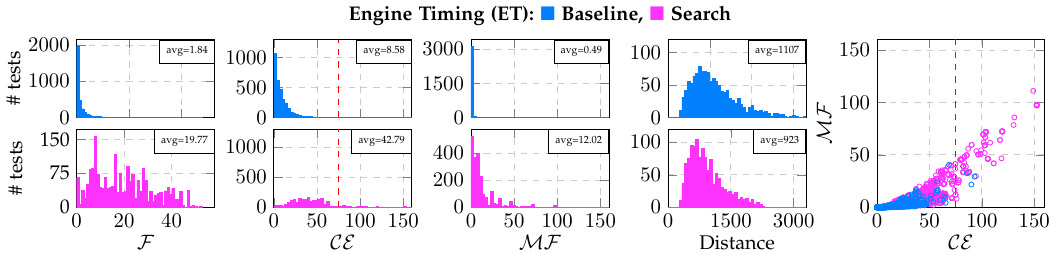}
    \includegraphics{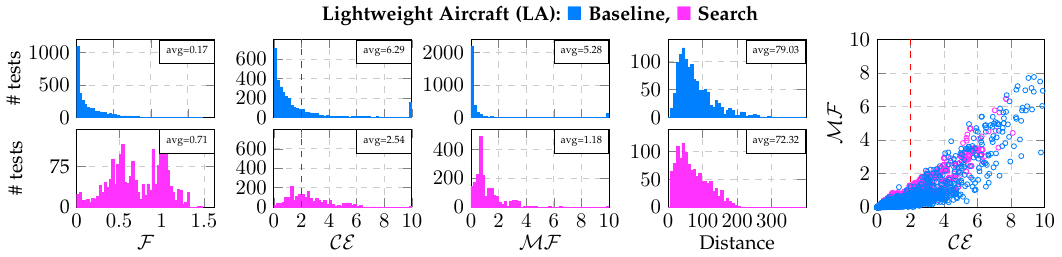}
    \caption{Results of the experimental campaign.
    The three groups of plots report the results for each test subject (CF, ET, and LA).
    In each group, the first row with the blue plots reports the results for the random baseline and the second row reports the results for the search.
    The first four columns report the histograms for the fitness $\fitness$, control error $\ctrlerr$, MR-falsification degree $\mrfalsify$, and distance.
    The scatter plots in the rightmost column show the control error and fitness for each test: the purple circles indicate tests obtained with the search while the blue circles represent tests obtained from the baseline.
    The red dashed lines indicate the control error threshold $\ctrlerrth$ used to distinguish the trivial failures.
    }
    \label{fig:empirical-results}
\end{figure*}

\begin{table}
    \centering
    \caption{Comparison with the Mann-Whitney U test of fitness, MR-falsification degree, and control error between search algorithm and baseline.
    The table reports both the \texttt{U} statistic and the \texttt{p} value for each test subject.
}
    \label{tab:base-vs-search-stat-tests}
    \def\lClmnLong {10mm}
\def\lClmnLongShort {7mm}
\newcolumntype{P}[1]{>{ \centering \arraybackslash }p{#1}}
\begin{tabular}{p{5mm}P{\lClmnLong}P{\lClmnLongShort}P{\lClmnLong}P{\lClmnLongShort}P{\lClmnLong}P{\lClmnLongShort}}
    \toprule
    \multicolumn{7}{c}{\bf Comparison of metrics distributions between search algorithm}\\
    \multicolumn{7}{c}{\bf and baseline with the Mann–Whitney U test.}\\ \toprule
     & \multicolumn{2}{c}{$\fitness$} &  \multicolumn{2}{c}{$\mrfalsify$} & \multicolumn{2}{c}{$\ctrlerr$}  \\ \cmidrule(lr){2-3}  \cmidrule(lr){4-5} \cmidrule(lr){6-7}
             & \texttt{U} & \texttt{p} & \texttt{U} & \texttt{p} & \texttt{U} & \texttt{p} \\ \cmidrule(lr){2-3}  \cmidrule(lr){4-5} \cmidrule(lr){6-7}
    {\bf CF} & $5.8\cdot10^{6}$ & 0.0 & $5.9\cdot10^{6}$ & 0.0 & $5.8\cdot10^{6}$ & 0.0 \\ \cmidrule(lr){2-3}  \cmidrule(lr){4-5} \cmidrule(lr){6-7}
    {\bf ET} & $6.2\cdot10^{6}$ & 0.0 & $6.2\cdot10^{6}$ & 0.0 & $6.2\cdot10^{6}$ & 0.0 \\ \cmidrule(lr){2-3}  \cmidrule(lr){4-5} \cmidrule(lr){6-7}
    {\bf LA} & $6.0\cdot10^{6}$ & 0.0 & $5.5\cdot10^{6}$ & 0.0 & $4.9\cdot10^{6}$ & 0.0 \\ \bottomrule
\end{tabular}
 \end{table}

\subsection{RQ1: Search Effectiveness}
\label{sec:rq-search}
\subsubsection{Methodology}
\label{sec:rq-search-methodology}
To assess the effectiveness of our search algorithm, we compare its results (the ones corresponding to the bold and purple values in Table~\ref{tab:parameters-mrs-investigation}) to the results of a baseline that relies on the random generation of programs using our proposed grammar.
To make the results comparable, we generate as many programs as the search generates across all generations.\footnote{
    This is based on the assumption that the time required for performing the computations associated with the search (like mutation, crossover, diversity computation) is negligible with respect the execution of the actual tests.
    This assumption applies to our test subjects and it is generally satisfied in CPSs, as they require the execution of physics simulators that are computationally demanding~\cite{Menghi:2019}.
}
This number is however not fixed and depends on how many individuals are selected for mutation and crossover at each generation.
Its maximum is the number of generations multiplied with $\lambda$, while to obtain the average we need to multiply this value by the probability of crossover and mutation.
In our case, the maximum corresponds to $40\cdot80=3200$ individuals, and the average to $40\cdot80\cdot0.7=2240$, where 0.7 is the sum of the mutually exclusive crossover and mutation rates.
For our baseline of comparison, we use the worst-case estimate of 3200 to avoid the risk of biasing the results in favour of the search. 

We aim to assess different aspects of the search, namely:
\begin{itemize}
    \item the effectiveness of the search in generating high-fitness individuals,
    \item the effectiveness of the search in generating effective test inputs, and
    \item the diversity of the high fitness individuals.
\end{itemize}
To assess the effectiveness of the search against the random generation of programs, we compare the obtained values for the fitness $\fitness$.
The distribution of the fitness values over all the generated programs is shown in the histograms in the first column of Figure~\ref{fig:empirical-results}.
In this figure, the different rows correspond to the three test subjects, with the blue histograms reporting the baseline results and the purple histograms reporting the search results.
For each histogram, we report the average in the top-right corner.
Importantly, to compare the distributions we use the Mann-Whitney U test, which allows us to assess the statistical significance of the difference between distributions.
The results of the test for the fitness distributions are reported in the $\fitness$ column of Table~\ref{tab:base-vs-search-stat-tests}.

To assess the effectiveness of the test inputs, since the fitness does not have a direct interpretation, we also compare the MR-falsification degree $\mrfalsify$ and control errors $\ctrlerr$.
As done in the case of the fitness, we compare the distribution of the two metrics values over all the generated programs with the histograms shown in the second and third columns of Figure~\ref{fig:empirical-results}; the results of the Mann-Whitney U test are listed in the $\mrfalsify$ and $\ctrlerr$ columns of Table~\ref{tab:base-vs-search-stat-tests}.
For CF and LA, $\mrfalsify$ and $\ctrlerr$ can reach very high values in some of the trivial failures, with the drone and aircraft flying very far from both reference and expected output.
For example, the drone flies up to tens of meters away from the reference and expected output, which is very large compared to the expected precision in the order of centimetres.
Such tests are not relevant for our analysis; thus, we limit the CF plots to values between 0 and 0.6 metres and the LA plots to values between 0 and 10 metres.\footnote{
    The tests with large $\ctrlerr$ and $\mrfalsify$ are counted in the right-most bin of each histogram.
    As shown in the plots, they constitute a small minority of the overall number of tests.
    The only exception is the LA baseline that generated a significant number of trivial failures.
}
In addition to the histograms, we study the relation between $\mrfalsify$ and $\ctrlerr$ for individual tests.
In fact, we are interested in tests that at the same time maximise the MR-falsification degree and satisfy the control error constraint.
To assess the two metrics in relation to one another, in the right-most column of Figure~\ref{fig:empirical-results}, we provide scatter plots of $\mrfalsify$ and $\ctrlerr$ where each observation is a test.
We include all the tests across generations for the search algorithm and all of the random tests generated for the baseline.
We also display a dashed red line indicating the control error threshold $\ctrlerrth$.
Since the objective of the search is to maximise $\mrfalsify$ while keeping $\ctrlerr$ below this threshold, {\em the relevant high fitness individuals are the ones to the left of the dashed line and further up in the scatter plots}.

Finally, to assess the diversity among high-fitness individuals, we also provide histograms of the Euclidean distances among the tests from the archive.
These values are shown, for each case study, in the fourth column of Figure~\ref{fig:empirical-results}.
By analysing these plots, we can check that high-fitness individuals are not all similar to each other, i.e., they capture diverse ways to test the CPS.

\subsubsection{Analysis}
\label{sec:rq-search-analysis}
The fitness histograms in Figure~\ref{fig:empirical-results} clearly show that the search generates tests with much higher fitness than the baseline: for CF the average fitness increases from 0.01 to 0.09, for ET from 1.84 to 19.77, and for LA from 0.17 to 0.71.
Comparing the $\ctrlerr$ and $\mrfalsify$ histograms, we observe that, when using the search, they also increase in value, indicating that our approach outperforms the baseline by achieving higher $\mrfalsify$ values at the cost of an acceptable $\ctrlerr$ increase.
All such increases are confirmed to be significant by the Mann-Whitney U test results in Table~\ref{tab:base-vs-search-stat-tests}.
The \texttt{U} statistic in the table quantifies how often observations from one distribution rank higher than those from the other distribution: the large values (all in the order of $10^6$) confirm the hypothesis that the search increases the three metrics.
The \texttt{p}-values are all numerically equivalent to 0, indicating that the obtained results are statistically significant.
These results show that maximising the MR-falsification degree while constraining the control error is not a simple problem that can be addressed by random generation.

From a diversity point of view, we observe in the distance histograms (the ones in the fourth column) small differences between the baseline and the search in all the three test subjects.
Since the baseline is purely exploratory, this confirms the effectiveness of the proposed mechanism for fostering diversity during the search (Section~\ref{sec:selection-mec-and-diversity}).

By comparing our three test subjects, we observe that they exhibit different testing challenges.
Looking at the scatter plots of CF and LA, we observe that the baseline is already able to generate a number of tests with both high $\ctrlerr$ (above the threshold) and $\mrfalsify$ values: such tests are the ones further up and right in the plot.
In contrast, for ET, the baseline generates mostly tests with low $\ctrlerr$ (below the threshold) and $\mrfalsify$ values.
This happens because it is easier to cause a failure in CF and LA, whereas ET is a more robust system.
Intuitively, when a drone (or plane) starts to misbehave, it may crash and irreparably fail, thus showing high $\ctrlerr$ and $\mrfalsify$ values.
An engine, instead, even after showing a deviation from the expected behaviour, is much more likely to recover and return to its nominal behaviour.

When it is easier to lead a system out of its design scope, like CF and LA, the challenge is to cause non-trivial failures rather than causing failures in general.
For such CPSs, our testing approach, using MR-falsification, enables the generation of more subtle failures, with low control error (below $\ctrlerrth$) but out of the design scope (i.e., unexpected behaviour).
Furthermore, it significantly reduces the number of trivially passing tests, i.e., the ones with low $\ctrlerr$ and $\mrfalsify$ values, thus reducing the resources needed for the testing campaign.
With a more robust system, like ET, simply causing failures is a challenge.
In this case, the search finds scenarios with higher MR-falsification than random generation, which necessarily also increases the control error---the tests are further up---but also further right in the scatter plot.
Our tests thus show that MR-falsification is also a valid fitness metric to guide test generation for ET, as it can find tests with higher control error but still avoids trivial fails.

{\bf Answer to RQ1.} Our experimental results indicate that our approach significantly outperforms the baseline, achieving higher MR-falsification degree and acceptable control error.
Specifically, we observe that satisfying the conflicting requirement of maximising the MR-falsification degree while constraining the control error is a challenging problem.
For CF, even though the baseline was also able to generate some failures, our approach increased the average $\mrfalsify$ values by more than two-fold (from 0.06 to 0.14), while keeping the average $\ctrlerr$ value at 0.16, thus very close to the threshold $\ctrlerrth=\qty{0.15}{\metre}$.
For LA, the baseline was also able to generate test cases with various values of $\ctrlerr$ and $\mrfalsify$.
However, the search significantly reduced the number of generated tests with low $\ctrlerr$ and $\mrfalsify$ values, thus making the testing campaign more efficient.
For ET, even though the system is robust and failing test cases are difficult to generate, the MR falsification was effective in guiding the search of test cases, not only with higher $\mrfalsify$ values but also higher $\ctrlerr$ values.
Notably, when compared to the baseline, the average $\ctrlerr$ value increased from 0.49 to 42.79, thus still well below the threshold $\ctrlerrth=\qty{75}{rpm}$.

\subsection{RQ2: Effectiveness of MRs as Oracle}
\label{sec:rq-oracle}
\subsubsection{Methodology}
\label{sec:rq-oracle-methodology}
To assess the ability of MR-falsification to improve oracles based on the control error, we analyse whether, among those that show control error values below the threshold $\ctrlerrth$ (thus excluding trivial fails), the metric helps distinguish passed and failed tests.
Specifically, we want to show that MR-falsification provides new information with respect to the control error.
We study the two metrics in relation to one another both in a  quantitative and in a qualitative way.

For the quantitative analysis, we use the R-squared statistic (also known as the coefficient of determination).
This metric measures how much variance in one variable can be explained by another variable.
R-squared is computed based on the residuals (i.e., the differences) that remain after performing a linear regression between $\mrfalsify$ and $\ctrlerr$.
It is normalised so that a value close to 1 indicates a strong relation between the two variables and 0 indicates no relation.
To give a graphical interpretation, the R-squared statistic assesses how much a linear regression can explain the datapoints on the left-hand side of the red dashed line (the control error threshold) in the scatter plots of Figure~\ref{fig:empirical-results}.
This allows us to quantify how strong the relation is between the two variables for tests with $\ctrlerr$ values below the threshold.

From a qualitative perspective, we analyse the $\mrfalsify$ and $\ctrlerr$ scatter plots in Figure~\ref{fig:empirical-results}.
We use them to observe trends in the relation between the two metrics.
Furthermore, we exemplify the effectiveness of the MR-falsification degree by reporting and analysing one high-fitness test from each of our test subjects.

\subsubsection{Analysis}
\label{sec:rq-oracle-analysis}
As part of the quantitative analysis, we measured a R-squared statistic of $0.48$ for CF, of $0.5$ for ET, and of $0.4$ for LA.
This means that the control error can explain approximately half of the variance that we observe in the MR-falsification degree.
The other half is then new information available to the engineers to distinguish between the passed and failed test cases.
This shows that the MR-falsification degree metric improves oracles based on the control error, by allowing engineers to distinguish test cases within and outside of the design scope.

In terms of the qualitative analysis, the rightmost plots of Figure~\ref{fig:empirical-results} graphically confirm that MR-falsification has higher variance for tests below the control error threshold (highlighted by the red dashed line).
The variation in $\mrfalsify$ values is clearly wider for the tests obtained using the search when compared to those obtained with the random baseline.
This is consistent with the answer to RQ1, which showed how the generation of non-trivial tests, which also falsify the MRs, requires a targeted search.
Thus, when the test generation does not explicitly target the MR-falsification, as in the baseline, fewer tests with different $\mrfalsify$ values are observed.

\begin{figure}
    \begin{subfigure}{\columnwidth}
        \includegraphics{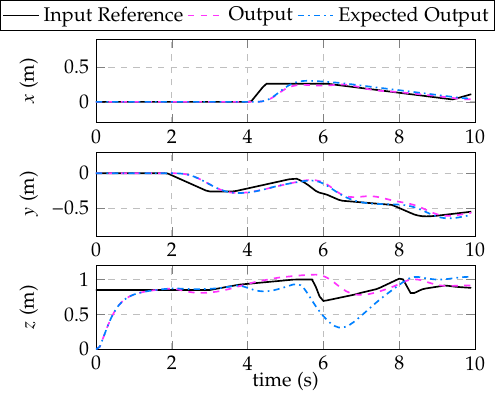}
        \caption{CF test subject.}
        \label{fig:cf-test-example}
    \end{subfigure}
    \vspace{4mm}

    \begin{subfigure}{\columnwidth}
        \includegraphics{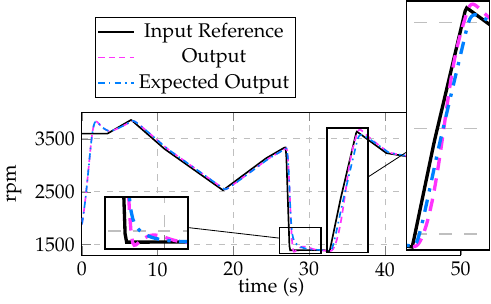}
        \caption{ET test subject.}
        \label{fig:et-test-example}
    \end{subfigure}
    \vspace{4mm}

    \begin{subfigure}{\columnwidth}
        \includegraphics{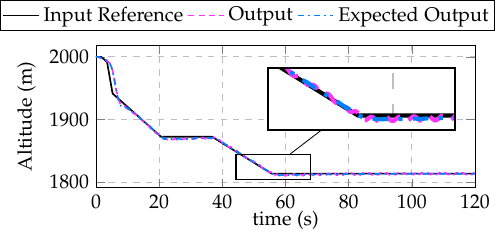}
        \caption{LA test subject.}
        \label{fig:la-test-example}
    \end{subfigure}
    \caption{Example of high-fitness tests for each test subject.}
    \label{fig:high-fitness-examples}
\end{figure}

When looking at the complete scatter plots, i.e., including the tests above the control error threshold, they seem to show that the MR-falsification is linearly correlated with the control error and thus does not provide new information.
This is expected for tests with larger control error values, since the MR-falsification degree and control error measure the deviation of the actual output from the expected output and the reference, respectively.
Since the expected output effectively tracks the reference (recall that the base test cases that are used to compute the expected output are designed to be within the design scope), when the actual output largely deviates from one of them, it will also deviate to a similar extent from the other.
However, when focusing on tests with control error values below the threshold, we observe more spread in $\mrfalsify$ values for the same $\ctrlerr$  value.
This confirms the weaker relation between the two variables, as we observed also with R-squared analysis.
Furthermore, we qualitatively observe that, below the threshold, the spread of $\mrfalsify$ values increases as the control error increases.
In other words, tests with very low (compared to the threshold) $\ctrlerr$  values tend to also have very low $\mrfalsify$ values.
This is also expected, as those tests are most likely well within the design scope, as the CPS effectively tracks the reference, thus leading to low $\mrfalsify$ values.

In Figure~\ref{fig:high-fitness-examples} we present examples of high-fitness individuals for each test subject.

Figure~\ref{fig:cf-test-example} shows such an example for CF; 
in the figure, we use the same notation as previous figures, with the black solid line showing the input reference, the blue dashed line showing the expected output computed based on the output of the initial test cases, and the purple dash-dotted line showing the output of the follow-up test case.
We also include three plots for the $x$, $y$ and $z$ dimensions in which the drone flies.
In this test of the CF, we observe a deviation between the actual and expected outputs along the $z$ direction between 4 and 8 seconds together with a smaller deviation along the other directions.
As we are familiar with the test subject, we are able to identify the root-cause for this failure: what we observe is an unexpected interaction between the altitude controller and the horizontal position controllers.
Specifically, the drone needs to tilt to move horizontally: in this way, the force generated by the propellers does not only push the drone straight upward but also pushes it in the horizontal direction.
However, in this scenario, this also reduces the vertical force and makes the drone lose altitude.
Differently, in other tests, this loss of altitude does not happen.
Thus, our test case generation is effectively generating an input sequence that causes this fault to appear.
Importantly, this faulty scenario can be automatically identified only thanks to the deviation between the expected and actual outputs: in fact, for this flight input sequence, traditional specifications like overshoot or steady-state error are not applicable.

For the ET, we show an example of generated failure in Figure~\ref{fig:et-test-example}, with the same graphical conventions as Figure~\ref{fig:cf-test-example}.
In the figure, we zoom-in into two subtle failures that our approach was able to generate and identify thanks to the MR-falsification degree.
Such subtle failures happen at around second 28 and 35 of the test.
For both failures, we observe that the actual output (in purple) overshoots compared to the expected output (in blue).
Notably, the actual output is even closer to the reference, thus showing a lower control error; however, the expected output is smoother and shows an expected behaviour that would put less stress on the engine.
Given that the actual output is still close to the input, in absence of the expected output (i.e., of the blue line), it would be difficult to have an automated approach that recognises such failures.
This shows how the expected output (and thus the MR-falsification degree) helps in identifying subtle failures in CPSs.

For the LA, we show an example of generated failure in Figure~\ref{fig:la-test-example} using the same conventions as for previous figures.
In this test, we observe a subtle oscillation in the actual output that was not predicted by the expected output.
This oscillation appears during most of the test, but we emphasise it with a zoom-in at around 60 seconds in the test.
Being relatively small, the oscillation does not increase the control error with respect to another test that does not show it.
However, our approach can detect it thanks to the deviation from the expected output.
Like for ET, this example shows how the MR-falsification degree provides new information to engineers to automatically identify failures in CPSs, which is specifically relevant for the identification of subtle failures.

The ET and LA failures show a deviation from the expected behaviour that is small in amplitude, justifying why we refer to them as subtle failures~\cite{zhang:2024,Hildebrandt:2020}.
However, depending on the application context, they can still have a high impact.
For example, in a car, the overshoot detected in Figure~\ref{fig:et-test-example} causes unwanted accelerations, which can lead to discomfort and, on the long run, wear of physical components.
In the LA test subject (Figure~\ref{fig:la-test-example}), an oscillation of the aircraft altitude, even if in the order of meters, is clearly undesirable, and it can not only cause discomfort, but also make the aircraft less robust to external perturbations like wind gusts.

{\bf Answer to RQ2.} Our tests show that MR-falsification effectively supports engineers in distinguishing test cases that show unpredictable behaviour of a CPS, though they are not trivially failing (acceptable control error).
The low R-squared value for all the three test subjects confirms that the MR-falsification degree is not correlated with the control error for test cases that do not trigger trivial failures.
Engineers have thus one more mechanism at their disposal, in addition to the control error, to identify tests that show unpredictable behaviour.
Furthermore, our tests show that this information is available in the range of control error values where it is most needed: when they are neither very low (i.e., tests that clearly show desirable behaviour of the CPSs) nor too high (i.e., tests that clearly show trivial failures of the CPSs).

\subsection{Threats to Validity}
In terms of internal validity, we identify a threat concerning the statistical analysis of the relation between the MR-falsification degree and the control error.
The relation analysis between these two metrics is limited by the impossibility of obtaining unbiased, sufficiently large samples of either of the quantities.
Specifically, we cannot generate input traces where we know a priori the control error (or the MR-falsification degree).
Thus, the analysis is limited by the fact that it is difficult to generate test cases with diverse $\mrfalsify$ values (as observed in RQ1).
Despite this, our statistical comparison results align with our findings based on visual inspection.
Our analysis shows that there is not a strong relation between the two quantities when considering the tests with $\ctrlerr$ values below the threshold, which is where we are interested in providing a new metric to distinguish between passed and failed test cases.

External validity concerns the generalisability of our results.
Although we successfully applied our approach to thee CPSs from the automotive, drone, and aircraft domains, there might be threats to the applicability of the approach to CPSs from other domains.
However, the fundamental component of our approach---the MRs--are based on the definition of a linear system, which is common to any CPS that is developed with the use of control theory and is independent of the specific application domain.
Thus, we contend that our approach is, in principle, applicable to any CPS that is developed leveraging control theory.

\subsection{Limitations}
The practical applicability of our approach might be limited by two main factors.

First, the need to tune the parameters of the fitness function: in our two test subjects, we needed to use different values for these parameters (Table~\ref{tab:parameters-mrs-investigation}).
When applying our approach to other CPSs, some iterations for choosing these values might be needed; this will increase the number of tests to be performed and thus the testing cost.
On the other hand, even a small campaign like the one we performed, based on standard values, was able to provide effective parameters.
This suggests that the tuning of the fitness function parameters should not be a major limitation.

Second, the proposed setup for the search requires a significant number of tests for the application of the approach (namely, around $40\cdot80\cdot0.7=2240$ tests).
This can be limiting for CPSs that require long execution time to test.
Reducing the execution time is outside the scope of our study.
However, for our experimental campaign, we did not perform any further optimisation of the search parameters after determining the values listed in Table~\ref{tab:parameters-gp}.
Therefore, it is likely possible to reduce the number of individuals or generations, thus reducing the number of CPS executions (tests) required by our approach.

\subsection{Data Availability}
\label{sec:data-availability}
The code implementing our approach and the test subjects are available in the following archives:
\begin{itemize}
    \item \url{https://doi.org/10.6084/m9.figshare.26303992}.
\end{itemize}
The data of the tests for the drone and engine are available here:
\begin{itemize}
    \item CF: \url{https://doi.org/10.6084/m9.figshare.26304028},
    \item ET: \url{https://doi.org/10.6084/m9.figshare.26303950},
    \item LA: \url{https://doi.org/10.6084/m9.figshare.28430612}.
\end{itemize}

Running the CF and ET experiments (including parameters assessment and random variability assessment) took around 60 hours for each test subject on a laptop with an Apple M1 Pro chip.
The LA test subject, instead, took around 96 hours on the Aion supercomputer of the University of Luxembourg.\footnote{
    The specs of the supercomputer can be found here: \url{https://hpc-docs.uni.lu/systems/aion/}.
    We estimated that running the LA test subject on a laptop would have required around 936 hours (39 days).
}

The significant difference in time between the test subjects is firstly due to the longer duration of the LA tests (see Table~\ref{tab:parameters-mrs-set}).
Intuitively, the flight of an aircraft takes much more time than the flight of a drone or the testing of an internal combustion engine.
Furthermore, the simulations of the aircraft use a model of the physics that is more complex than the one of the other test subjects.
This complexity results in more computational time needed for the CPS execution.
 \section{Related Work}
\label{sec:related-work}
The oracle problem for CPSs is not new in the software engineering literature.
Several specification languages and associated runtime verification tools have been proposed to specify CPSs requirements as oracles~\cite{Boufaied:2021}.
\citet{Kapinski:2016} provide a library of common CPSs behaviours, including steady-state error and overshoot (mentioned in Section~\ref{sec:introduction}); notably, the authors also provide templates for expressing such properties in STL~\cite{Furia:2012}.
\citet{Boufaied:2021} proposed a taxonomy of CPSs properties (like the overshoot mentioned in Section~\ref{sec:introduction}) and reviewed the expressiveness of various signal-based temporal logics in terms of the property types identified in the taxonomy.
\citet{Menghi:2021} as well as \citet{Dawes:2022} proposed specification languages for expressing hybrid properties of CPSs, capturing both the software and the physical behaviour.
Several approaches have proven the practical applicability of specification-driven runtime verification tools to the oracle problem in the CPSs domain, both in offline~\cite{Menghi:2021,Boufaied:2020, db:fse2024} (in the aerospace domain) and online settings~\cite{Kane:2014} (in the automotive domain).
Our approach, differently from these works, does not discuss the satisfaction of the requirements, but rather the satisfaction of the design assumptions expressed as MRs.

Few works on CPSs testing use concepts from control engineering.
The most closely related work is from \citet{Hildebrandt:2020}, as the authors seek to generate subtle (i.e., non-trivial) failures in trajectories of mobile robots (i.e., in the jargon of this article, sequences of input position references).
The authors use models of the physics (the same ones used during the development of control algorithms) to compute the set of feasible input traces and then use user-defined stress metrics to guide the test-generation process.
In this work, we instead use the MR-falsification degree as a new (stress) metric and the control error threshold to target non-trivial failures.
This makes our approach more general and application independent.
\citet{He:2019} used system identification~\cite{bittanti:2019} (a sub-field of control engineering) to implement oracles for CPSs output traces.
Other works connect to control engineering aspects of CPSs by expanding temporal logics with frequency-domain operators~\cite{Nguyen:2017, Donze:2012}.
The frequency domain is widely used in traditional control engineering as alternative to the time domain to represent systems and signals, as it is well-suited to analyse equation-based models of the physical world.
While such works move toward bridging the gap between software and control engineering in CPSs, none of them integrates the guarantees that can be obtained from the two disciplines.
One of the main strengths of our approach is the use of the design assumptions to achieve such synergy.

CPSs requirements expressed in signal-temporal logic have also been used to guide the generation of CPSs test cases~\cite{zhang:2024}.
Such works fall under the class of CPSs-falsification techniques.
Notably, \citet{zhang:2024} discuss the relevance of finding subtle requirements violations.
Among falsification approaches, we highlight two prior works that leverage control engineering tools, and more specifically system identification~\cite{bittanti:2019}.
In such works, system identification techniques are used to reduce the number of tests that have to be executed in order to find faults~\cite{Menghi:2019} or to reduce the parameter definition effort required by genetic algorithms~\cite{Aleti:2015}.
However, these approaches, while they do leverage control theoretical tools, they still target the falsification of the requirements.
Differently, we use the control theoretical models to define a metric (the MR-falsification degree) to guide the test case generation, as part of the testing objective itself.

Very early work in the metamorphic testing (MT) literature investigated its applicability to CPSs, specifically control systems~\cite{Chen:2011}.
More recent work applied MT to an elevator system, using MRs concerning the system performance and quality of service~\cite{Ayerdi:2020,Ayerdi:2021,Ayerdi:2022}.
\citet{Li:2021} performed a preliminary exploration of MT applicability to multi-module UAVs, using hand-written MRs.
\citet{Deng:2022} applied MT to autonomous driving using MRs derived from traffic rules and domain knowledge.
Finally, \citet{ayerdi:2023} used MRs to perform runtime verification of autonomous driving systems.
In contrast, our work clearly departs from the above works by defining MRs using the design assumptions, instead of requirements satisfaction.
Notably, by using assumptions that underlie any control-theoretical model, we define MRs that are applicable to any CPS from any domain, as long as it is developed based on control theory.
 \section{Conclusions}
\label{sec:conclusions}
We addressed the problem of generating CPSs input traces with potentially arbitrary shapes together with associated oracles.
Since CPSs requirements are defined only for simple input traces, we instead used the concept of design assumptions: when the design assumptions hold, engineers can rely on the control theoretical guarantees on the CPSs performance.
Specifically, we used the linear behaviour design assumption to define MRs that can be used for the generation of new input traces with associated expected output traces.
We then set out to search for input traces that falsify the MRs while avoiding trivial failures.
To perform this search, we defined a grammar for programs representing the possible combinations of MR applications.
Using this grammar, our GP-based approach searches for input traces that falsify the MRs while avoiding large control error (associated with trivial failures).

We evaluated our testing approach on three test subjects: a drone, an engine, and an lightweight aircraft.
Our results show that falsifying the MRs while avoiding trivial failures is a complex problem.
Moreover, they show that MR falsification complements the use of the sole control error for identifying passed and failed test cases.
Specifically, in test cases where the control error is not decisive (i.e., test cases where the control error value is not large enough to clearly identify a failure), MR falsification shows different values.
Such a difference allows engineers to distinguish test cases that do or do not satisfy the design assumptions.

In future work, we plan to expand the approach to also generate external environmental inputs that can affect the CPS, like wind gusts for a drone or road slopes in a car.
Such inputs need to be handled differently as the CPS is expected to minimise their impact on its behaviour rather than to track them.
Another interesting direction is to use the MRs to address the test case generation in combination with other testing objectives.
Furthermore, we plan to address the testing of other design assumptions listed in our previous work~\cite{mandrioli:2023b}, such as those concerning the interaction with discrete inputs and the validity of the physics models.
 
\section*{Acknowledgments}
This project has received funding from SES; the Luxembourg National Research Fund under the Industrial Partnership Block Grant (IPBG), ref. IPBG19/14016225/INSTRUCT; and the European Union’s Horizon Europe program under Grant Agreement No.\ 101148870 (ContTestCPS).
Lionel Briand was partly funded by the Research Ireland grant 13/RC/2094-2 and NSERC of Canada under the Discovery and CRC programs.
For the purpose of open access, and in fulfillment of the obligations arising from the grant agreements, the authors have applied a Creative Commons Attribution 4.0 International (CC BY 4.0) license to any Author Accepted Manuscript version arising from this submission. 
The experiments presented in this paper were carried out using the HPC facilities of the University of Luxembourg~\cite{HPCCT22} (see \texttt{\href{http://hpc.uni.lu}{hpc.uni.lu}}).

\bibliographystyle{IEEEtranN}
\bibliography{main}

\end{document}